\documentclass[fleqn,usenatbib]{mnras}
\usepackage{newtxtext,newtxmath}
\usepackage{pdflscape}	
\usepackage[T1]{fontenc}
\usepackage{ae,aecompl}
\usepackage{rotating}
\DeclareRobustCommand{\VAN}[3]{#2}
\let\VANthebibliography\thebibliography
\def\thebibliography{\DeclareRobustCommand{\VAN}[3]{##3}\VANthebibliography}

\usepackage{graphicx}	
\usepackage{amsmath}	

\title[Binary Fraction of RSGs in the LMC and SMC]{The Binary Fraction of Red Supergiants in the Magellanic Clouds}

\author[Min Dai et al.]{
Min Dai,$^{1,2}$
Shu Wang$^{3}$\thanks{E-mail: shuwang@nao.cas.cn}
and Biwei Jiang$^{1,2}$\thanks{E-mail: bjiang@bnu.edu.cn}
\\
$^{1}$School of Physics and Astronomy, Beijing Normal University, Beijing 100875, People's Republic of China\\
$^{2}$Institute for Frontiers in Astronomy and Astrophysics, Beijing Normal University, Beijing 102206, People's Republic of China\\
$^{3}$CAS Key Laboratory of Optical Astronomy, National Astronomical Observatories, Chinese Academy of Sciences, Beijing 100101, People's Republic of China
}

\date{Accepted 2025 April 03. Received 2025 April 03; in original form 2024 October 23}

\pubyear{2024}

\begin{document}
\label{firstpage}
\pagerange{\pageref{firstpage}--\pageref{lastpage}}
\maketitle

\begin{abstract}
	Red supergiants (RSGs), as the descendants of OB-type stars and the progenitors of supernovae, provide crucial insights into the evolution of massive stars, particularly in binary systems. Previous studies show that the binary fraction of RSGs ($\approx 15\% - 40\%$) is significantly lower than that of their predecessors ($\approx 50\% - 70\%$). In this work, we investigate the binary fraction of RSGs with the recently selected largest samples of 4695 and 2097 RSGs in the Large Magellanic Cloud (LMC) and Small Magellanic Cloud (SMC), respectively. The binary system with a hot companion (O-, B- and A-type star) is identified by detecting the ultraviolet (UV) excess in the observed spectral energy distribution (SED) ranging from ultraviolet to mid-infrared after subtracting the model SED of RSG since RSGs are very weak in the UV band. It is found that the lower limit of binarity is 30.2\% $\pm$ 0.7\% and 32.2\% $\pm$ 1\% in the LMC and SMC, respectively. If the sample is limited to luminous RSGs with log $L/L_{\odot} > 4.0$, the binary fraction becomes 26.6\% $\pm$ 1.1\% and 26.4\% $\pm$ 1.7\% in the LMC and SMC, respectively. The derived binary fraction is valid in the range of $\sim$ 2.3 < $\log P / [\text{d}]$ < $\sim$ 8. Our study suggests that roughly one-third of massive stars host a third companion within $\sim$ 30,000 AU. In addition, 15 RSGs are also identified as binary via HST/STIS spectra, and a handful of the binaries identified by the SED fitting are confirmed by their light curve and radial velocity dispersion. The stellar parameters of the companions, i.e. $T_{\mathrm{eff}}$, $R$, $L$ and log $g$, are calculated by model fitting.
\end{abstract}




\section{Introduction} \label{sec:intro}

Red supergiants (RSGs) are massive stars located in the upper right of the Hertzsprung-Russell (H-R) diagram. The initial mass of RSGs is about between $6.1 - 30$ $M_{\odot}$ \citep{2024ApJ...965..106Y}. As the evolved descendants of OB-type main-sequence stars, RSGs have low effective temperatures ($T_{\mathrm{eff}} \approx 3500 - 4500$ K), very high luminosities ($L/L_{\odot} \approx 4 - 5.8$) and large radii ($R \approx 1500$ $R_{\odot}$) \citep{2005ApJ...628..973L,2007ApJ...660..301M,2008IAUS..250...97M,2012A&A...540L..12W,2013ApJ...767....3D,2016ApJ...826..224M}.

Most massive stars are found in binary or in multiple star system \citep{1998AJ....115..821M,2001A&A...368..122G,2012Sci...337..444S,2014ApJS..215...15S,2014ApJS..213...34K}. The binary fraction of OB-type stars in the period range $0 \lesssim \log P / [\text{d}]  \lesssim 3$ was determined to be at least $\sim 50\% - 70\%$ \citep{2008ASPC..387...93G,2012Sci...337..444S,2013A&A...550A.107S,2015A&A...580A..93D,2022A&A...665A.148S,2024arXiv241016114S}. However, previous studies found the binary fraction of RSGs $\approx$ 15\% to 40\%, employing diverse methodologies that correspond to detecting different orbital period ranges, which will be discussed in subsequent sections \citep{2019AA...624A.129P,2020AA...635A..29P,2020ApJ...900..118N,2021ApJ...908...87N,2021MNRAS.502.4890D,2022MNRAS.513.5847P}, significantly lower than their progenitors. RSGs are not only descendants of OB-type stars, but also the progenitors of certain types of supernovae. RSGs with masses between about 8 and 25 $M_{\odot}$ will explode as Type II-P supernovae, as predicted by single star models \citep{2006ApJ...637..914W,2013A&A...558A.131G}. The evolution of RSG in a binary system can be classified into two distinct categories. If the orbital period is greater than 1,500 days \citep[very wide binary systems,][]{2012Sci...337..444S}, the evolution is identical to that of a single star. However, many Type II supernova progenitors undergo mass exchange before exploding \citep{2019A&A...631A...5Z}, and interacting binaries may be the key to understanding the diversity of Type II supernova light curves \citep{2018PASA...35...49E}. The study of RSG binary populations can contribute to the development of evolutionary models of massive stars. 

Two methods are generally used to study the binary fraction of RSGs. One way is to analyze the variability of radial velocity (RV). In this way, \citet{2019AA...624A.129P} derive that the upper limit of binary fraction of RSGs is 30\% in a sample of 17 RSGs in the 30 Doradus region of the Large Magellanic Cloud (LMC), with the orbital period range $3.3 \leq \log P / [\text{d}]  \leq 4.3$. Later, \citet{2020AA...635A..29P} report a binary fraction of $30 \pm 10\%$ in a sample of 15 RSGs in the NGC 330 region of the Small Magellanic Cloud (SMC), with the orbital period range $2.3 \leq \log P / [\text{d}]  \leq  4.3$. \citet{2021MNRAS.502.4890D} find a minimum binary fraction of $14 \pm 5 \%$ in a sample of 51 cool supergiants (CSGs) in the LMC and of $15 \pm 4 \%$ from a sample of 72 CSGs in the SMC, with the orbital period range $2.8 \lesssim \log P / [\text{d}]  \lesssim 3.6$. The other way is to detect the blue-star spectral features (BSSFs) from spectra (e.g., \citeauthor{2018AJ....156..225N} \citeyear{2018AJ....156..225N}). \citet{2020ApJ...900..118N} further develop the BSSFs method. They use a machine-learning method based on $k$-nearest-neighbor algorithm (k-NN) with $U$, $B$, $V$, and $I$ photometry to classify stars as single or binary. They find the binary fraction of RSGs = $19.5_{-6.7}^{+7.5}\%$ with a sample of 1820 RSGs in the LMC. The period sensitivity range by the single-epoch spectroscopy and photometric identification is not explicitly mentioned in the respective articles. Afterwards, \citet{2021ApJ...908...87N} apply a similar method to a sample of 1909 and 1702 RSGs in M31 and M33, respectively. They find that the binary fraction of RSGs in M33 shows a strong dependence on galactocentric distance ($41.2_{-7.3}^{+12}\%$ at inner regions; $15.9_{-1.9}^{+8.6}\%$ at outer regions) and the binary fraction of RSGs = $33.5_{-5.0}^{+8.6}\%$ in M31. 

In addition to the two methods mentioned above, \citet{2022MNRAS.513.5847P} make use of the fact that the brightness of RSGs is significantly fainter than that of main-sequence B-type stars in the near- and far-UV (NUV and FUV, respectively) to distinguish binary RSGs from single. They report a binary fraction of $18.8 \pm 1.5 \%$ in a sample of 560 RSGs in the SMC by using the Ultraviolet Imaging Telescope (UVIT) FUV (0.172 $\mu$m) photometry, with the orbital period range $3 \leq \log P / [\text{d}]  \leq 8$. In brief, a RSG is regarded as in a binary once detected by UVIT at 0.172 $\mu$m. Based on the same rule that the detection of UV excess is an indicator of early-type star companion, this work attempts to identify the binary RSGs in the LMC and SMC by fitting the spectral energy distribution (SED) of RSGs in the MCs. While RV method exhibits constraints in its detectable orbital period range (e.g., $\log P / [\text{d}]  \leq 4.3$), previous photometric studies have been limited by incomplete datasets. We therefore employ multi-band photometric data to investigate the binary fraction of RSGs in the MCs. The UV photometry is taken from several observations including SMASH, HST, GALEX, $Swift$/UVOT, XMM-OM and SMSS besides UVIT. Furthermore, the sample of RSGs is recently renewed significantly based on the astrometric measurements and new method to exclude the foreground dwarfs stars. From the SED fitting, we are able to identify the RSG binary with hot companions (e.g. A-, B-, O-type stars).
The principle idea of this method is that the fluxes in the red and near-infrared (NIR) bands is dominated by the RSG, while in the UV bands, it is dominated by the hot companion.

This paper is structured as follows. In Section \ref{sec:data_samples}, we describe the datasets for constructing SEDs and the sample of RSGs in the MCs. The methods for obtaining the binary fraction of RSGs and the parameters of both RSGs and their companions are presented in Section \ref{sec:Method}. In Section \ref{sec:Discussions}, we compare our results with previous works, the distributions of our samples in the color-color diagrams, extra evidence of identifying binaries, and the limitations of this study. Finally, the main conclusions are presented in Section \ref{subsec:summary}.


\section{Data and Sample} \label{sec:data_samples}
\subsection{Data}\label{datasets}

The photometric data from UV to infrared (IR) are collected to construct the SED. The surveys used include: (1) X-ray Multi-Mirror Mission Optical Monitor \citep[XMM-OM,][]{2001A&A...365L...1J,2001A&A...365L..36M}, (2) UltraViolet and Optical Telescope \citep[UVOT,][]{2004AIPC..727..637G,2005SSRv..120...95R}, (3) UV Imaging Telescope \citep[UVIT,][]{2012SPIE.8443E..1NK,2014Ap&SS.354..143H}, (4) Hubble Space Telescope (HST), (5) Galaxy Evolution Explorer \citep[GALEX,][]{2005ApJ...619L...7M,2007ApJS..173..682M}, (6) the Survey of the Magellanic Stellar History \citep[SMASH,][]{2017AJ....154..199N}, (7) the SkyMapper Southern Survey \citep[SMSS,][]{2018PASA...35...10W}, (8) Gaia \citep{2023A&A...674A...1G}, (9) the Two Micron All Sky Survey \citep[2MASS,][]{2006AJ....131.1163S},(10) the Wide-field Infrared Survey Explorer \citep[WISE,][]{2010AJ....140.1868W}, and (11) Spitzer Space Telescope \citep[Spitzer,][]{2004ApJS..154....1W}. All filters information used in this work are presented in Table \ref{passbandinfo}. Besides, the spectroscopic data are collected from the Apache Point Observatory Galactic Evolution Experiment \citep[APOGEE,][]{2017AJ....154...94M}, Gaia low-resolution blue/red prism photometer BP/RP mean spectra \citep[XP spectra,][]{2023A&A...674A...3M} and the Hubble Space Telescope Imaging Spectrograph (HST/STIS). The brief introduction to these projects are following.

\subsubsection{Photometry} \label{subsec:Photometry}
XMM-OM is the optical monitor of the $XMM$-$Newton$ space observatory. It provides photometries in UVW2 (0.2 $\mu$m), UVM2 (0.23 $\mu$m), UVW1 (0.29 $\mu$m), $U$, $B$ and $V$ \citep{2001A&A...365L...1J,2001A&A...365L..36M}. The PSF FWHM aperture size is $\sim$ 2 $''$ in the UV bands. The photometric data are taken from XMM-OM serendipitous UV source survey catalogue 5.0 \citep{2012MNRAS.426..903P}.

$Swift$/UVOT is a telescope of $Swift$ mission. It provides photometries in UVW2 (0.21 $\mu$m), UVM2 (0.22 $\mu$m), UVW1 (0.28 $\mu$m), $U$, $B$ and $V$ \citep{2004AIPC..727..637G,2005SSRv..120...95R}. The PSF FWHM aperture size is $\sim$ 0.9 $''$ at 0.35 $\mu$m. The photometric data are taken from the serendipitous UV source catalogs \citep{2014Ap&SS.354...97Y}.

UVIT is a telescope of ASTROSAT mission. It provides 17 bands for FUV ($130-180$ nm), NUV ($200-300$ nm) and blue-visible ($320-550$ nm) \citep{2012SPIE.8443E..1NK,2014Ap&SS.354..143H,2014SPIE.9144E..1SS}. The PSF FWHM aperture size is $\sim$ 1.40$''$ in F172M. \citep{2024ApJS..275...34P} Part of the FUV and NUV photometric data are taken from \citet{2022MNRAS.513.5847P} and \citet{2023ApJ...946...65D}.

HST is a 2.4-m space telescope. The UV ($\lambda < 0.38$ $\mu$m) photometric data are taken from Hubble Source Catalog (HSC \footnote{\url{https://catalogs.mast.stsci.edu/hsc}}). The PSF FWHM aperture size of WFC3/UVIS is $\sim$ 0.075$''$ in the UV bands. The criterion of ``Flags = 0'' is adopted to ensure the data quality \citep{2016AJ....151..134W}.

$GALEX$ is an all-sky survey using a 50-cm telescope to provide NUV (0.23 $\mu$m) and FUV (0.15 $\mu$m) photometry \citep{2005ApJ...619L...7M,2007ApJS..173..682M}. The photometric data are taken from the revised catalog of GALEX ultraviolet sources (GUVcat$\_$AIS, \citealt{2017ApJS..230...24B}). The spatial resolutions (FWHM) are 4.3$''$ and 5.3$''$ in the FUV and NUV, respectively. The criteria of ``largeobjsize = 0'', 13.73 $<$ FUV $<$ 19.9 mag and 13.85 $<$ NUV $<$ 20.8 mag are adopted to ensure the data quality.

SMASH is a survey project by using the dark energy camera on the NOAO Blanco 4-m telescope at the Cerro Tololo Inter-American Observatory (CTIO) to observe the Magellanic Clouds and their periphery. It observes about 480 square degrees of sky with a depth down to ~24th mag in $u$, $g$, $r$, $i$, $z$ \citep{2017AJ....154..199N}. The PSF FWHM aperture size is $\sim$ 1$''$ in $u$ and $g$ bands. The $u$ and $g$ band photometric data are taken from data release 2 (DR2) \citep{2021AJ....161...74N}, and the criterion of photometric error $<$ 0.05 mag is adopted to ensure the data quality.

SMSS is a southern sky survey by using the 1.3-m SkyMapper telescope at Siding Spring Observatory in Australia in six bands, i.e. $u$, $v$, $g$, $r$, $i$, $z$ \citep{2018PASA...35...10W}. The photometric accuracy of the DR2 is 1\% in $u$ and $v$, and 0.7\% in $griz$ \citep{2019PASA...36...33O}. The median PSF FWHM aperture size is 3.1$''$, 2.9$''$, 2.6$''$, 2.4$''$, 2.3$''$ and 2.3$''$ in the $u$, $v$, $g$, $r$, $i$ and $z$ bands. The criteria of ``x\_flags = 0'' and ``x\_nimaflags = 0'' are adopted to ensure the data quality. Besides, the filters of SMSS $u$ and SMASH $u$ are similar. For the sources overlapped with SMASH $u$, a linear fit is performed between SMSS $u$ and SMASH $u$ by using the RANSAC algorithm of the scikit-learn package. The observations of SMSS $u$ and $v$ with 1$\sigma$ outliers are removed since the SMSS $u$ band has a brighter limiting magnitude and some known issues described by \citet{2019PASA...36...33O}.

The third Gaia data release, Gaia DR3, provides broadband photometry in the $G$, $G_{\mathrm{BP}}$ and $G_{\mathrm{RP}}$ bands for more than 1.5 billion sources, radial velocity (RV) measurements for more than 33 million bright stars, and low-resolution blue/red prism photometer BP/RP mean spectra for about 220 million sources \citep{2023A&A...674A...1G}. The spatial resolutions of Gaia is $\sim$ 1.5$''$\citep{2021A&A...649A...1G}. The $G_{\rm BP}$ and $G_{\rm RP}$ brightness are used in consucting SEDs, and the RV data are used in excluding foreground dwarf stars and identifying binaries.

2MASS is an all-sky survey that provides photometry in the NIR $J$, $H$, and $K_{\mathrm{S}}$ bands. Its 10$\sigma$ point-source detection depths are better than 15.8, 15.1, and 14.3 mag in the three bands, respectively \citep{2006AJ....131.1163S}. The median PSF FWHM aperture size is $\sim $3$''$ in the $J$, $H$, and $K_{\mathrm{S}}$ bands. The $J$, $H$, and $K_{\mathrm{S}}$ data are taken from 2MASS all-sky point-source catalog \citep{2003yCat.2246....0C}.

WISE is an all-sky survey in four infrared bands in W1, W2, W3 and W4 \citep{2010AJ....140.1868W}. The angular resolution is 6.1$''$, 6.4$''$, 6.5$''$ and 12.0$''$ in the W1, W2, W3 and W4 bands. The photometric data are taken from the AllWISE catalog \citep{2014yCat.2328....0C}.

Spitzer is an infrared space telescope. It carried three infrared instruments, the Infrared Array Camera (IRAC), the Infrared Spectrograph (IRS), and the MIPS \citep{2004ApJS..154....1W}. The spatial resolution is $\sim$ 2$''$ in IRAC bands. The four IRAC bands and [24] band of MIPS data are taken from Spitzer Science Center \& Infrared Science Archive \citeyearpar{2021yCat.2368....0S}.

\subsubsection{Spectroscopic Data} \label{subsec:spectrum}
APOGEE is a spectroscopic survey of the Sloan Digital Sky Survey (SDSS) project. It produces the high-resolution (R $\sim$ 22,500)  infrared (1.51–-1.70 $\mu$m) spectra by using the 2.5-m Sloan Telescope and du Pont Telescope \citep{2006AJ....131.2332G,2017AJ....154...94M,2019PASP..131e5001W}. The stellar parameters ($T_\mathrm{eff}$ and RV) are taken from DR17 \citep{2022ApJS..259...35A}.

The Gaia XP spectra are a low-resolution ($\lambda/\Delta\lambda$ = 30 $-$ 100) data product released by Gaia DR3 \citep{2021A&A...652A..86C}. The XP spectra covers the 330 nm to 1050 nm wavelength range. The XP spectra are used only in the range of 400 to 900 nm because of systematic effects at $\lambda < 400$ nm and $\lambda > 900$ nm \citep{2023A&A...674A...3M}. The XP spectra of possibly poor quality are omitted by using a 3$\sigma$ criterion, where the $\sigma$ is calculated by comparing the photometry of $G_\mathrm{BP}$ and $G_\mathrm{RP}$ from Gaia DR3 with synthetic photometry of $G_\mathrm{BP}$ and $G_\mathrm{RP}$ derived from XP spectra. The calibrated spectra are obtained by using GaiaXPy \footnote{\url{https://gaia-dpci.github.io/GaiaXPy-website/}} \citep{daniela_ruz_mieres_2023_8239995}. In the process of fitting the companion SED, the XP spectra are used to replace the photometry in the UVOT/$BV$ bands, XMM-OM/$BV$ bands, SMASH/$g$ band, SMSS/$gri$ bands and APASS all bands.

HST spectra are collected from Space Telescope Imaging Spectrograph (STIS). The HST/STIS grating spectrum used in this work is named G230LB, and it covers the wavelength from 0.17 $\mu$m to 0.31 $\mu$m, and has a spectroscopic resolution of R $\sim$ 700 \citep{medallon2023stis}.

\subsection{The RSG Samples} \label{sec:Samples}

Until 2020, only a few hundred RSGs have been identified in the LMC and SMC \citep{10.1093/mnras/193.2.377,10.1093/mnras/197.2.385,1983ApJ...272...99W,10.1046/j.1365-8711.2000.03196.x,Massey_2002,Massey_2003,Yang_2011,Yang_2012,Neugent_2012_,2015A&A...578A...3G}. With updated data and methods for identifying RSGs in the MCs, the number of RSGs has increased by orders of magnitude in recent years. \citet{2020A&A...639A.116Y,2021A&A...646A.141Y} identify 2974 (LMC) and 1239 (SMC) RSGs by using an astrometric solution from Gaia/DR2 and multiband color–magnitude diagrams (CMDs). \citet{2023ApJ...946...43W} construct a spectroscopic sample with 1073 (LMC) and 398 (SMC) RSGs by using astrometric data from Gaia and spectroscopic data from APOGEE. \citet{2020ApJ...900..118N} report 4090 RSGs with log $L/L_{\odot}$ $>$ 3.5 in the LMC by using astrometric data from Gaia/DR2 and CMD. \citet{Massey_2021} identify 1745 RSGs in the SMC by using the same method as \citet{2020ApJ...900..118N}.

We use the sample of \citet{2021ApJ...923..232R} which includes 4823 and 2138 RSGs in the LMC and SMC, respectively. They use the updated astrometric data (Gaia/EDR3) and intrinsic NIR color–color diagram (CDD) to remove foreground dwarfs, and identify RSGs in the intrinsic CMD. The sample of \citet{2021ApJ...923..232R} contains 77\% (LMC) and 74\% (SMC) of the RSGs from \citet{2020A&A...639A.116Y,2021A&A...646A.141Y}, 93\% (LMC) and 95\% (SMC) of the RSGs from \citet{2023ApJ...946...43W}, 95\% (LMC) of the RSGs from \citet{2020ApJ...900..118N}, and 91\% (SMC) of the RSGs from \citet{Massey_2021}. Compared to previous samples, this one is larger in number or purer. In order to further purify the sample, 128 (LMC) and 41 (SMC) foreground dwarfs are removed by adopting the criteria with RV $< 200$ km s$^{-1}$ for the LMC and RV $< 95$ km s$^{-1}$ for the SMC, where the RV values are taken from the APPOGE or Gaia catalog. Finally, a sample of 4695 and 2097 RSGs in the LMC and SMC is used in further analysis.

A radius of 1$''$ is adopted to cross-match the RSG sample with photometric and spectroscopic catalogs listed in Section \ref{datasets}. The number of stars cross-matched with each catalog is as follows:
\begin{itemize}
	\item For SMASH data, 6265 RSGs (92.24\%) in the $u$ band.
	\item For XMM-OM data, 71 (1.05\%), 90 (1.32\%), 641 (9.43\%) and 106 (1.56\%) RSGs in the UVW2, UVM2, UVW1 and $U$ band, respectively.
	\item For $Swift$/UVOT data, 32 (0.47\%), 8 (0.12\%), 46 (0.68\%) and 83 (1.22\%) RSGs in the UVW2, UVM2, UVW1 band $U$ band, respectively.
	\item For UVIT data, 87 (1.28\%) and 5 RSGs (0.07\%) in the F172M and N279N band, 4 RSGs (0.06\%) in the F154W and F169M band, 3 RSGs (0.04\%) in the N219M, N245M and N263M band, respectively.
	\item For HST data, 1 (0.01\%), 18 (0.27\%), 22 (0.32\%), 2 (0.03\%), 29 (0.43\%), 57 (0.84\%), 6 (0.09\%) and 3 (0.04\%) RSGs in the F170W, F225W, F275W, F280N, F300W, F336W, F343N and F373N band, respectively.
	\item For GALEX data, 5 RSGs (0.07\%) in the FUV and NUV band.
	\item For SMSS data, 347 (5.11\%), 811 (11.94\%), 3217 (47.36\%) RSGs in the $u$, $v$ and $z$ band.
	\item For Gaia data, 6755 (99.46\%) and 6756 (99.47\%) RSGs in the $G_{\rm{BP}}$ and $G_{\rm{RP}}$ band.
	\item For 2MASS data, 6792 RSGs (100\%) in the $J$, $H$ and $K_{\rm{S}}$ band.
\end{itemize}

\section{Method and Result} \label{sec:Method}

\subsection{Fitting the SED of RSGs} \label{subsec:RSGs}

The observational SED of RSG is fitted to retrieve the stellar parameters and form the basis to judge whether the UV excess exists. The model spectra are taken from the synthetic stellar spectra of Lejeune et al. (\citeyear{1997A&AS..125..229L}; hereafter L97). The L97 library covers the $T_\mathrm{eff}$ from 2500 K to 50,000 K, log $g$ from -1.02 to 5.0, and [M/H] from -3.5 dex to 1.0 dex. In order to improve the fitting precision, the interval of $T_\mathrm{eff}$ is narrowed from 250 K down to 50 K by interpolation.

The interstellar extinction correction is necessary to the photometric results at this step. Due to the red colors of RSGs, the effective wavelength $\lambda\mathrm{_{eff}}$ of the photometric band can be quite different from the claimed one generally for an A0-type star, in particular in the UV bands. Thus $\lambda\mathrm{_{eff}}$ is first calculated at the typical stellar parameters of RSGs, specifically $T_\mathrm{eff}$= 4000 K, log $g$ = 0, and [M/H] = -0.5 for the LMC, [M/H] = -1.0 for the SMC,
\begin{equation}
	\lambda_\mathrm{eff} \equiv \frac{\int \lambda T(\lambda) S(\lambda) \, d\lambda}{\int T(\lambda) S(\lambda) \, d\lambda}\text{,}
	\label{eq:effective_wavelengths}
\end{equation}
where $T(\lambda)$ is the filter transmission, and $S(\lambda)$ is the stellar model spectrum. To obtain $A\mathrm{_{\lambda}}$, $E(V - I)$ is taken from the OGLE extinction map by \citet{2021ApJS..252...23S}, and then converted to $A\mathrm{_{\lambda}}$ with the extinction law by \citet{2023ApJ...946...43W}. The fitting is later performed to the SED after extinction correction.

Since the flux \text{at} the wavelength shorter than 0.75 $\mu$m may be affected by the companion and the flux \text{at} the wavelength longer than 2.5 $\mu$m may be affected by circumstellar dust, the SED of RSG component is fitted only to the observations at the wavelength between 0.75 $\mu$m to 2.5 $\mu$m, including Gaia $G_{\mathrm{RP}}$, SMSS $z$, and 2MASS $J$, $H$, $K_{\mathrm{S}}$. To assess the goodness of fitting, the reduced $\chi^2$ for each star is calculated as
\begin{equation}
\begin{split}
 \frac{\chi^2}{\mathrm{dof}}=\frac{\sum_{j=1}^{N_{\mathrm{obs}}}[F_\nu(\lambda_j)^{\mathrm{mod}}-F_\nu(\lambda_j)^{\mathrm{obs}}]^2/w(\lambda_j)^2}{N_{\mathrm{obs}}-N_{\mathrm{para}}}\text{,}
\end{split}
\label{eq:chi-square}
\end{equation}
 where $F_\nu(\lambda_j)^{\mathrm{mod}}$ and $F_\nu(\lambda_j)^{\mathrm{obs}}$ are the model and the observed flux after correcting the extinction, respectively. $F_\nu(\lambda_j)^{\mathrm{mod}}$ is calculated by
\begin{equation}
F_\nu(\lambda_j)^{\mathrm{mod}} \equiv \frac{\int T(\lambda) S(\lambda) \, d\lambda}{\int T(\lambda) \, d\lambda}\text{.}
\label{eq:mod_flux}
\end{equation}
The weight is defined as $w(\lambda_j) = [F_\nu(\lambda_j)^{\mathrm{mod}} + F_\nu(\lambda_j)^{\mathrm{obs}}]/2$ to eliminate the effect caused by \text{the} large flux difference between various bands. $N_\mathrm{obs}$ is the number of observational data and $N_\mathrm{para}$ is the model parameter. Indeed, only $T_\mathrm{eff}$ is fitted as log $g$ and [M/H] are set to be constant for one galaxy, i.e. log $g$ = 0\footnote{Due to model limitations, log $g$ = 0.28 for $T_\mathrm{eff}$ < 3500 K.}, [M/H] = -0.5 (LMC) or [M/H] = -1.0 (SMC). The influence of various parameters will be discussed later and it's small. With the $T_\mathrm{eff}$  fitted, stellar radius ($R$) and luminosity ($L$) are calculated by
\begin{equation}
    R = d \sqrt{\frac{F_\nu(\lambda)^{\mathrm{obs}}}{F_\nu(\lambda)^{\mathrm{mod}}}}\text{,}
\label{eq:cal_R}
\end{equation}
\begin{equation}
	L = 4\pi R^2 F\text{,}
	\label{eq:cal_L}
\end{equation}
where $d$ is the distance of the MCs, and $F$ is the integrated flux of model spectrum. The distance modulus = 18.49 mag for the LMC and 18.94 mag for the SMC are adopted to calculate the $d$ \citep{2014AJ....147..122D,2015ApJ...815...87C}. The stellar parameters derived from the SED fitting are summarized in Table \ref{tab:Parameters of RSGs and companions}.

\begin{figure*}
	   \centering
	   \vspace{-0.0in}
	   \includegraphics[angle=0,width=85mm]{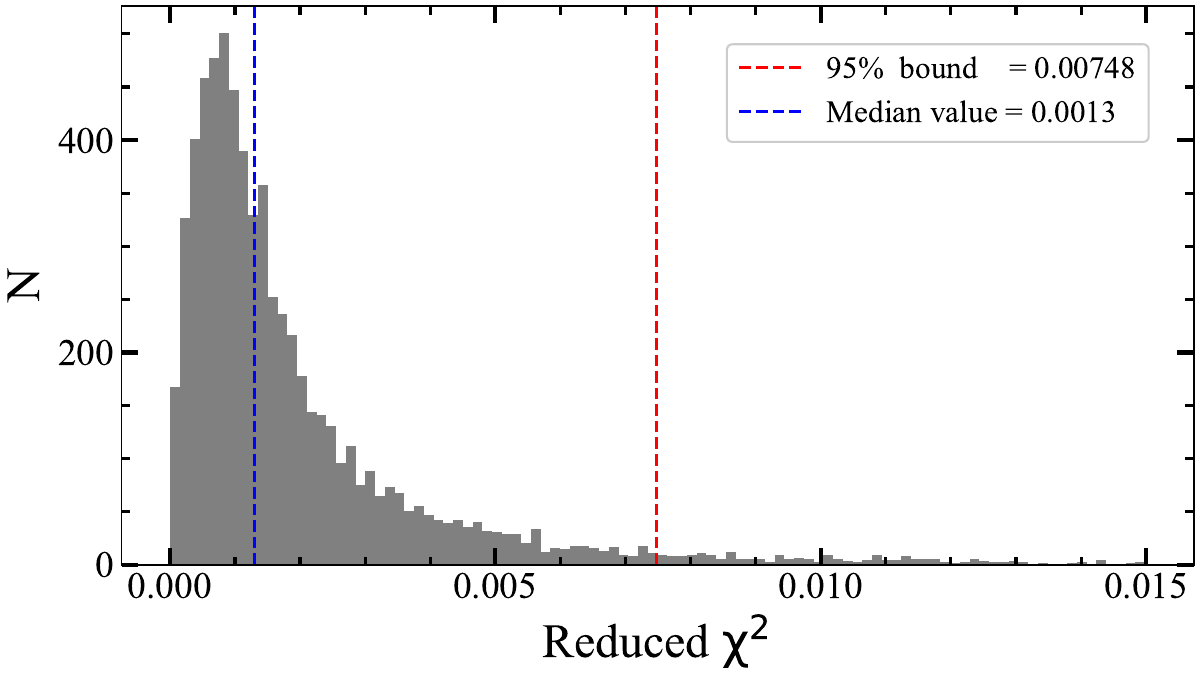}
	   \vspace{-0.0in}
	   \caption{\label{RSG_chi_distribution}The reduced $\chi^{2}$ distribution derived from the SED fitting of RSGs. The red dashed line and the blue dashed line represent the 95\% confidence interval and the median value of our sample, respectively.}
	   \end{figure*}

\begin{figure*}
		  \centering
		  \vspace{-0.0in}
		  \includegraphics[angle=0,width=170mm]{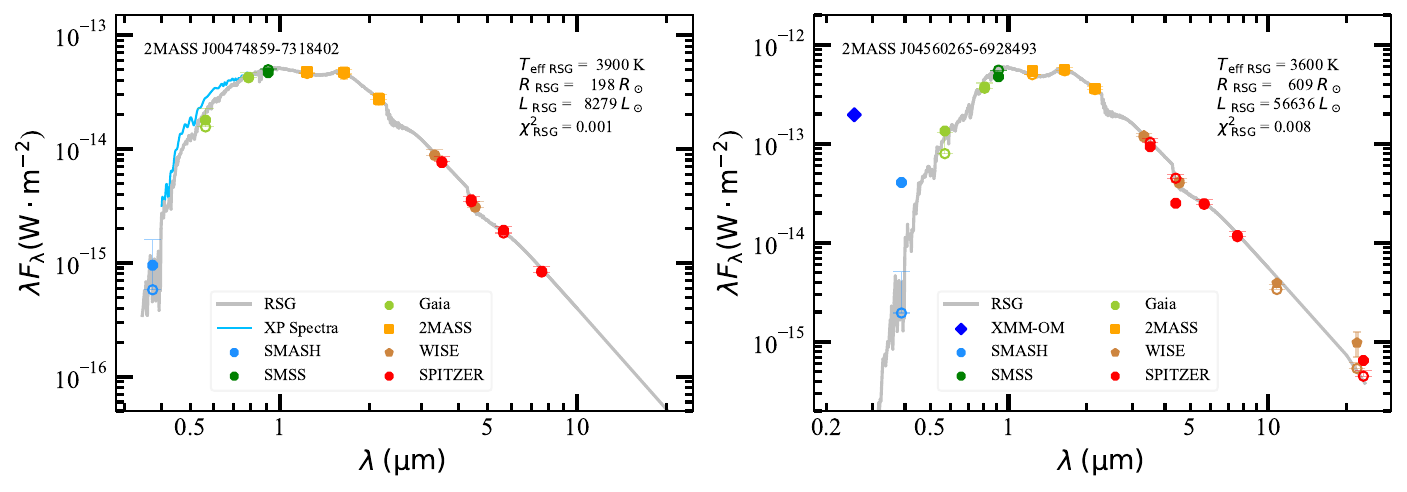}
		  \vspace{-0.0in}
		  \caption{\label{large_chi_square_and_UV_excess} The examples of the RSG SED fitting for 2MASS J00474859-7318402 and J04560265-6928493. The left panel shows a single RSG SED with a reduced $\chi^2$ around the median value, and the right panel shows a binary SED with a reduced $\chi^2$ greater than the 95\% confidence value. The gray solid line represents the L97 model spectrum, and the blue solid line represents the XP spectra. The solid and hollow markers with 3$\sigma$ error bars show the photometry and model data. The coordinate is marked on the upper left, and $T_{\mathrm{eff}}$, $R$, $L$ and $\chi^2$ are displayed on the upper right. The symbol conversions are followed in the following figures.}
	   \end{figure*}

Figure \ref{RSG_chi_distribution} shows the distribution of the reduced $\chi^2$ for fitting the SED of RSGs. The red dashed line in Figure \ref{RSG_chi_distribution} denotes the critical value of $\chi^2$ = 0.00748 at the 95\% confidence level. A reserved sample of 4390 and 2063 RSGs in the LMC and SMC is selected by removing the objects with $\chi^2 >$ 0.00748. It is worth noting that we check the SED fitting diagrams of all stars with reduced $\chi^2 >$ 0.00748 (305 in the LMC, 34 in the SMC), and find that 14\% of them show significant excess UV flux. This means that a small number of RSG binaries may be missed in this work. Two examples are displayed in Figure \ref{large_chi_square_and_UV_excess}. The left panel is the result of a single RSG star with a reduced $\chi^2$ around the median value, while the right panel is that of a candidate binary RSG with a reduced $\chi^2$ greater than the 95\% confidence value.

\begin{figure*}
		  \centering
		  \vspace{-0.0in}
		  \includegraphics[angle=0,width=85mm]{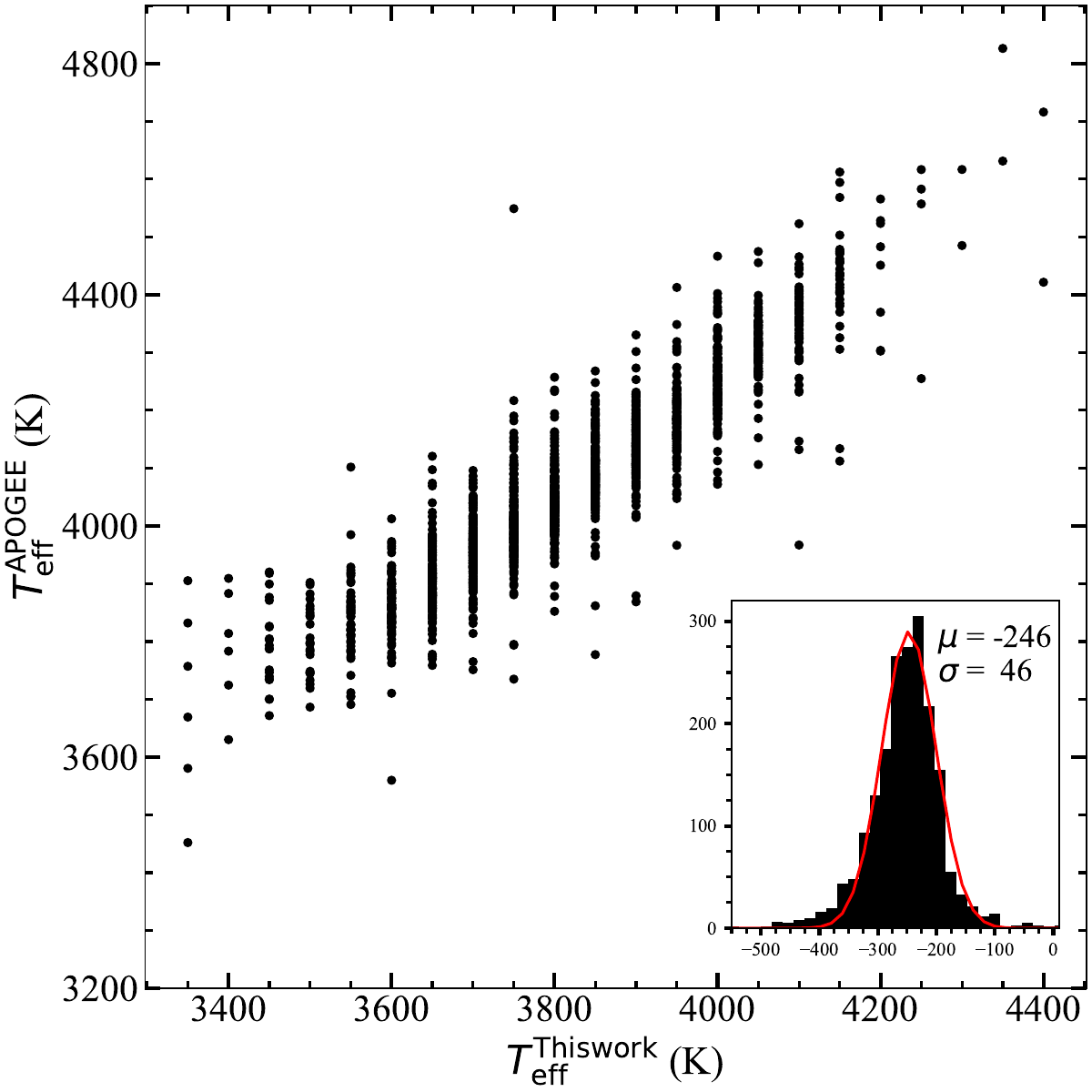}
		  \vspace{-0.0in}
		  \caption{\label{comparison_of_this_work_and_APOGEE_Teff}A comparison of $T_\mathrm{eff}$ from APOGEE and  from the SED fitting. The $\mu$ and $\sigma$ of residual are presented in the inset.}
	   \end{figure*}

The typical uncertainty of stellar parameters is estimated by the MCMC method for one star, 2MASS J04494147-6904402, whose photometric errors in $J$, $H$, $K_{\rm{S}}$ are around the median value of all sample stars. A total of 5000 SED fitting is performed to the observational SED that are generated from a gaussian distribution with photometric flux and error. The yielded errors of stellar parameters are $\sigma_{T_\mathrm{eff}} = 50$ K, $\frac{\sigma_R}{R} = ^{+1.4} _{-2.2} \%$ and $\frac{\sigma_L}{L} = ^{+1.36}_{-1.37} \%$. This result is considered as typical errors of stellar parameters for all RSGs. Although the $T_\mathrm{eff}$ we derived exhibit a systematic discrepancy compared to those provided by APOGEE (see Figure \ref{comparison_of_this_work_and_APOGEE_Teff}), the dispersion in their distribution remains at $\sim$ 50 K. Furthermore, the uncertainties in the derived $R$ and $L$ are relatively small, as these parameters were determined through SED convolution. The errors include the observational and model fitting, but do not consider the distance error.

Among these sources, 1950 are observed by APOGEE, whose $T_\mathrm{eff}$ from the SED fitting is compared with that from the APOGEE high resolution spectrum. As shown in Figure \ref{comparison_of_this_work_and_APOGEE_Teff}, there is an overall consistency between them, which confirms the method. Meanwhile, $T_\mathrm{eff}$ from the SED fitting is systematically lower than that from APOGEE by 246 K, and the difference has a dispersion of about 46 K. This can be attributed to the difference in stellar model and photometric uncertainty. Furthermore, the atmospheric parameters derived in our study exhibits excellent consistency with the spectroscopic determinations reported by \citet{2020ApJ...900..118N}.

\subsection{Identification of RSG Binary} \label{subsec:Identification_Binary}

Whether a RSG is a binary is justified by comparing  $F'_\nu(\lambda_j)^{\mathrm{mod}}$ with $F'_\nu(\lambda_j)^{\mathrm{obs}}$. $F'_\nu(\lambda_j)^{\mathrm{mod}}$ is the recalculated model flux according to individual $T_\mathrm{eff}$ from the SED fitting for each RSG, and $F'_\nu(\lambda_j)^{\mathrm{obs}}$ is the recalculated observed flux after correcting interstellar extinction based on the newly calculated $\lambda'_{\mathrm{eff}}$ for $T_\mathrm{eff}$. However, the transmissions of some UV filters \footnote{F225W, F170W bands (HST), UVM2, UVW1, UVW2 bands (XMM-OT), UVM2, UVW2, UVW1 bands ($Swift$/UVOT).} have a long tail at long wavelength side, causing the calculated $\lambda_{\mathrm{eff}}$ close to maximum $\lambda$ of the filter. For these affected filters, we cut off the long tails by removing the minimum 10\% of transmission. In addition, the transmission of UVIT filter is unavailable, the mean wavelength of filter is adopted as $\lambda_{\mathrm{eff}}$ \footnote{\url{https://uvit.iiap.res.in/Instrument/Filters}}.

By analyzing the L97 model spectra, we find that the larger log $g$ of RSG, the \text{higher} UV model fluxes, but the effect on the NIR wavelengths is small. In addition, the literatures suggest that the log $g$ of RSG can be up to 0.5 \citep{2017ars..book.....L}. Considering the effects from log $g$ and $T_\mathrm{eff}$, the error of modeled flux is
\begin{equation}
	F_{\nu}^{\mathrm{err}}(\lambda_j)^{\mathrm{mod}} = F_\nu^{\mathrm{max}}(\lambda_j)^{\mathrm{mod}} - F_\nu(\lambda_j)^{\mathrm{mod}}  \text{,}
\end{equation}
where $F_{\nu}^{\mathrm{err}}(\lambda_j)^{\mathrm{mod}}$ is the error of the model flux, $F_\nu^{\mathrm{max}}(\lambda_j)^{\mathrm{mod}}$ is the model fluxes with log $g$ = 0.5\footnote{Due to model limitations, log $g$ = 0.6 for $T_\mathrm{eff}$ < 3500 K.} and $T_{\rm eff}$ = $T_{\rm eff}^{\mathrm{RSG}}$ $+$ 50 K. $T_{\rm eff}^{\mathrm{RSG}}$ and $F_\nu(\lambda_j)^{\mathrm{mod}}$ are the $T_{\rm eff}$ and model flux of RSGs from SED fitting (see Section \ref{subsec:RSGs}).

Our binary identification criterion is expressed as
\begin{equation}
F_\nu(\lambda_j)^{\mathrm{obs}} - F_\nu(\lambda_j)^{\mathrm{mod}}> 3 F^{{\mathrm{err}}}_{\text{\(\nu\)}}(\lambda_j)^{\mathrm{obs}}  + 3 F^{{\mathrm{err}}}_{\text{\(\nu\)}}(\lambda_j)^{\mathrm{mod}} \text{,}
\label{eq:binary_criterion}
\end{equation}
where $F^{{\mathrm{err}}}_{\text{\(\nu\)}}(\lambda_j)^{\mathrm{obs}}$ is the error of observations. The RSG with photometric data at $\lambda_j < 0.4$ $\mu$m that is satisfied with Equation \ref{eq:binary_criterion} will be identified as a binary. In other words, the star is identified as a binary when the observational flux exceeds the model by three sigma in the UV band. An example of RSG binary SED is shown in Figure \ref{RSG SED fitting}, where the observed fluxes at $\lambda_\mathrm{eff} <$ 0.4 $\mu$m are significantly higher than the RSG model spectrum (gray solid line), and it is identified as a binary with a hot companion.

With the criterion described by Equation \ref{eq:binary_criterion}, 1325 and 664 binaries are identified in the LMC and SMC, respectively. The observed binary fraction is then 30.2\% $\pm$ 0.7\% (1325/4390) in the LMC and 32.2\% $\pm$ 1\% (664/2063) in the SMC and summarized in Table \ref{tab:Reported binary fraction of RSGs}. Here, the error in fraction is derived from the binomial distribution. In addition, due to the actual physical separation between the RSG and its putative companion may be remarkably large, the orbital period can reach to $\log P / [\text{d}]$ $\sim$ 8 (see discussion in section \ref{subsec:Range_Periods}). Under conditions of extreme orbital separation, the companion may not only lack significant gravitational interaction with the primary star, but may not be a real binary system. The main results together with the binary fraction of RSGs by other works (see Section \ref{subsec:Comparison}) are listed in Table \ref{tab:Reported binary fraction of RSGs}. 

\subsection{Stellar Parameters of the Companion} \label{subsec:Companion}

Stellar parameters of the companion star for the RSG binaries are further determined by fitting the residual SED after subtracting the RSG model SED from the observed SED. In agreement with the criteria to select hot companion, the wavelengths with $\lambda_\mathrm{eff}<$1 $\mu$m are used to fit the SED of companion. Before fitting the SED of the companion, the $statsmodels$ package is used to perform the locally weighted scatterplot smoothing (LOWESS) algorithm to smooth the Gaia/XP, HST/STIS and L97 spectrum. Then, the XP and STIS spectra are used in the process of fitting the companion SED. For unknown reasons, the XP spectra in the range of 600 to 900 nm reduce the goodness of fit, so only the XP spectra from 400 to 600 nm are used. In addition, the PAdova and TRieste Stellar Evolution Code (PARSEC) evolutionary tracks are used to confine the stellar parameter space of companions, which is illustrated in Figure \ref{hot_range}. In Figure \ref{hot_range}, black circles represent the values of $T_\mathrm{eff}$ and log $g$ which are adopted in the SED fitting of the companion. The reduced $\chi^2$ is again calculated by Equation \ref{eq:chi-square}.

One example of the companion SED fitting is shown in Figure \ref{SED fitting}, which shows the best fit of the SED for both the hot companion and RSG (left panel), as well as their positions in the H-R diagram (right panel). The $T_\mathrm{eff}$, $R$, $L$ and log $g$ derived from the SED fitting are marked as well. Compared to RSGs, the stellar parameters of companions are less accurate because many binaries lack the FUV photometry or with large photometric uncertainty. When fitting the parameters of companions, the varying numbers of UV photometric bands ($\lambda$ < $\sim$ 0.4 $\mu$m) range from 1 to 8. We present typical $T_\mathrm{eff}$ uncertainties of companions for $T_\mathrm{eff}$ ranges corresponding to 3 UV photometric bands (as objects with more than 3 UV bands are fewer). We employed 3,000 MCMC iterations to estimate the uncertainties, which systematically incorporate both $T_\mathrm{eff}$ of RSGs and photometric measurement errors. The derived median $T_\mathrm{eff}$ uncertainties for companion stars are 5.8\% in the 7,500 - 10,000 K range (73 stars), 5.5\% in the 13,000 - 18,000 K range (31 stars) and 6.2\% in the 20,000 - 30,000 K range (10 stars).

Figure \ref{HR_of_companion} shows the H-R diagram of all companion stars and their corresponding RSGs. The gray connecting lines delineate the corresponding associations between individual RSGs and their respective companion stars. The companions are mostly the main-sequence stars with $T_\mathrm{eff}$ ranging from 6000\,K to 30,000\,K. The limit at the low end, 6000K, indicates the lower limit of temperature that the UV excess can diagnose. It can be seen that the reduced $\chi^2$ is larger for lower-temperature star because their UV excess is less significant. The higher the temperature, the fewer the number of companions. This agrees with the general distribution. In addition, a few stars are on the horizontal branch, which can be understood by their relatively high temperature. Previous studies suggest that the companions of RSGs are O-,B-,A-type main-sequence stars and most of them are B-type \citep{1996A&A...307..829B,2018AJ....156..225N,2019ApJ...875..124N}. However, in this work, we find that 56\% of the companions (1107/1989) have $T_\mathrm{eff} < 7400$\,K. The reason is that the effective temperature and radius of a star are coupled, and when the number of available UV bands is small during the SED fitting, the result tends to be at the lower temperature. After eliminating the companions with $T_\mathrm{eff} < 7400$\,K, the distribution of the remaining companions by temperature is as follows:
\begin{itemize}
	\item 4.5\% (40/882) companions have $T_\mathrm{eff} \geq 25,000$\,K, corresponding to O-type star.
	\item 46.3\% (408/882) companions have $ 10,000 \leq T_\mathrm{eff} < 25,000$\,K, corresponding to B-type star.
	\item 49.2\% (434/882) companions have $ 7400 \leq T_\mathrm{eff} < 10,000$\,K, corresponding to A-type star.
\end{itemize}
Approximately half of the companions belong to the A-type class. These `cool' companions remained undetected in previous surveys. Figure \ref{SED fitting_for_cool} presents the SED fitting diagram of 2MASS J01045611-7215070 with $T_\mathrm{eff}$ = 8250 K as an illustrative example. While the flux at $\lambda$ $\approx 0.38$ $\mu$m shows no significant excess above the RSG model predictions, a pronounced flux excess becomes evident at $\lambda$ $\leq$ $\sim$ 0.38 $\mu$m compared to the RSG model expectations.
\citet{2019ApJ...875..124N} report a sample of 24 RSG+B binaries which are confirmed by spectrum in the LMC and SMC. Our sample has 20 common stars with theirs, and 18 among them are identified as binaries in this work. Such high coincidence (90\%) also indicates that the RSG star with hot companion can be identified successfully by the SED fitting method. On the other side, only 11 companions are also identified as B-type star. By checking all the SED fitting diagrams of other stars, it is found that these stars have few or no UV data except the band with $\lambda\mathrm{_{eff}} \approx 0.38$ $\mu$m, e.g. SMASH/$u$, SMSS/$u,v$, XMM-OM/$U$ and UVOT/$U$. This means that the stellar parameters of companions derived from the SED fitting using only the $\lambda\mathrm{_{eff}} \approx 0.38$ $\mu$m photometric band are not highly reliable and may incline to lower effective temperature. In other words, the RSG binaries can be well identified by the SED fitting based on current UV photometric data, but there is some uncertainty in stellar parameters of the companions, in particular the mid-type stars. The stellar parameters of companions are summarized in Table \ref{tab:Parameters of RSGs and companions}.

\section{Discussion} \label{sec:Discussions}

\subsection{Identification with HST Spectra} \label{subsec:HST_Spectra}
\begin{figure*}
		  \centering
		  \vspace{-0.0in}
		  \includegraphics[angle=0,width=170mm]{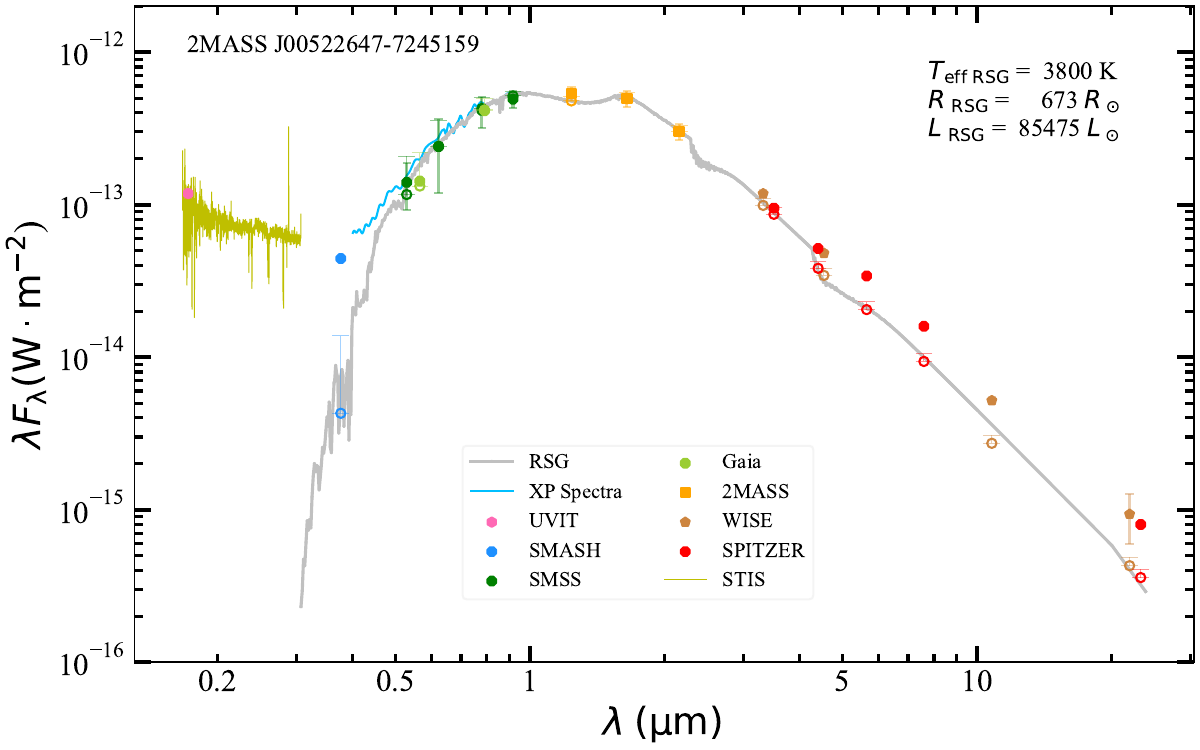}
		  \vspace{-0.0in}
		  \caption{\label{RSG SED fitting}An example of the SED fitting to the RSG component of a binary candidate, 2MASS J00522647-7245159. }
	   \end{figure*}

\begin{figure*}
	   \centering
	   \vspace{-0.0in}
	   \includegraphics[angle=0,width=85mm]{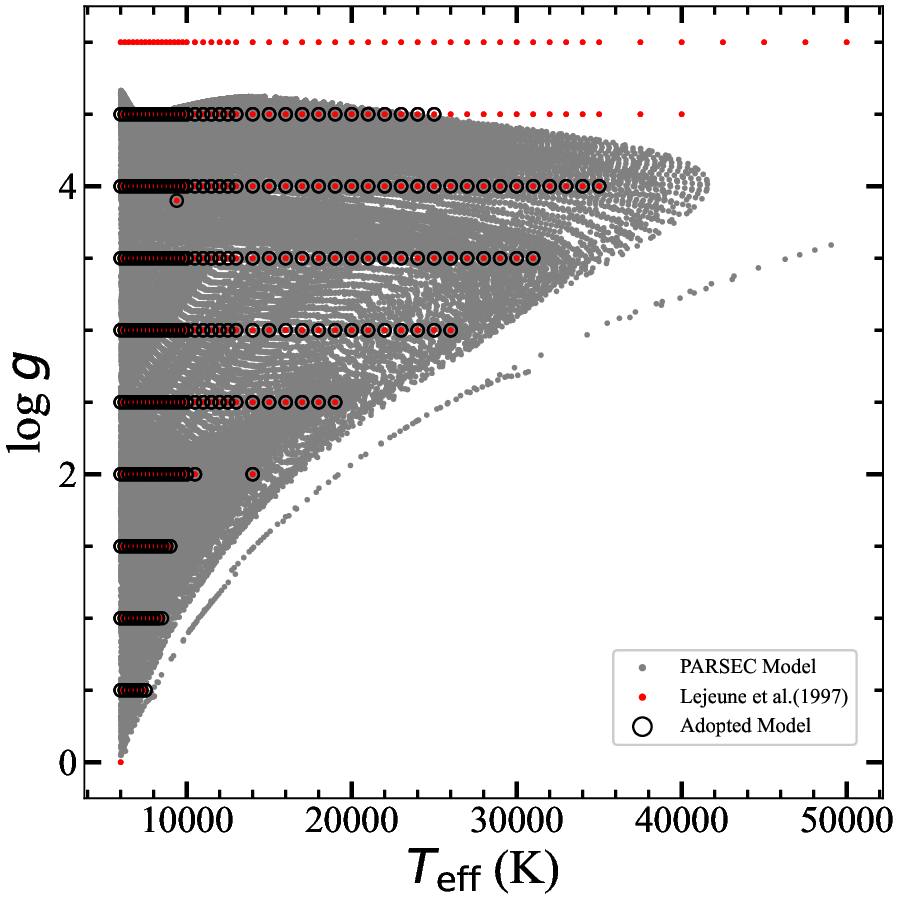}
	   \vspace{-0.0in}
	   \caption{\label{hot_range}The parameters space ($T_{\mathrm{eff}}$ vs. log $g$) for fitting the companion star. The red and gray solid dots represent the L97 and PARSEC model, respectively. The black circles denote the adopted stellar parameters in this work.}
	   \end{figure*}

\begin{figure*}
	   \centering
	   \vspace{-0.0in}
	   \includegraphics[angle=0,width=170mm]{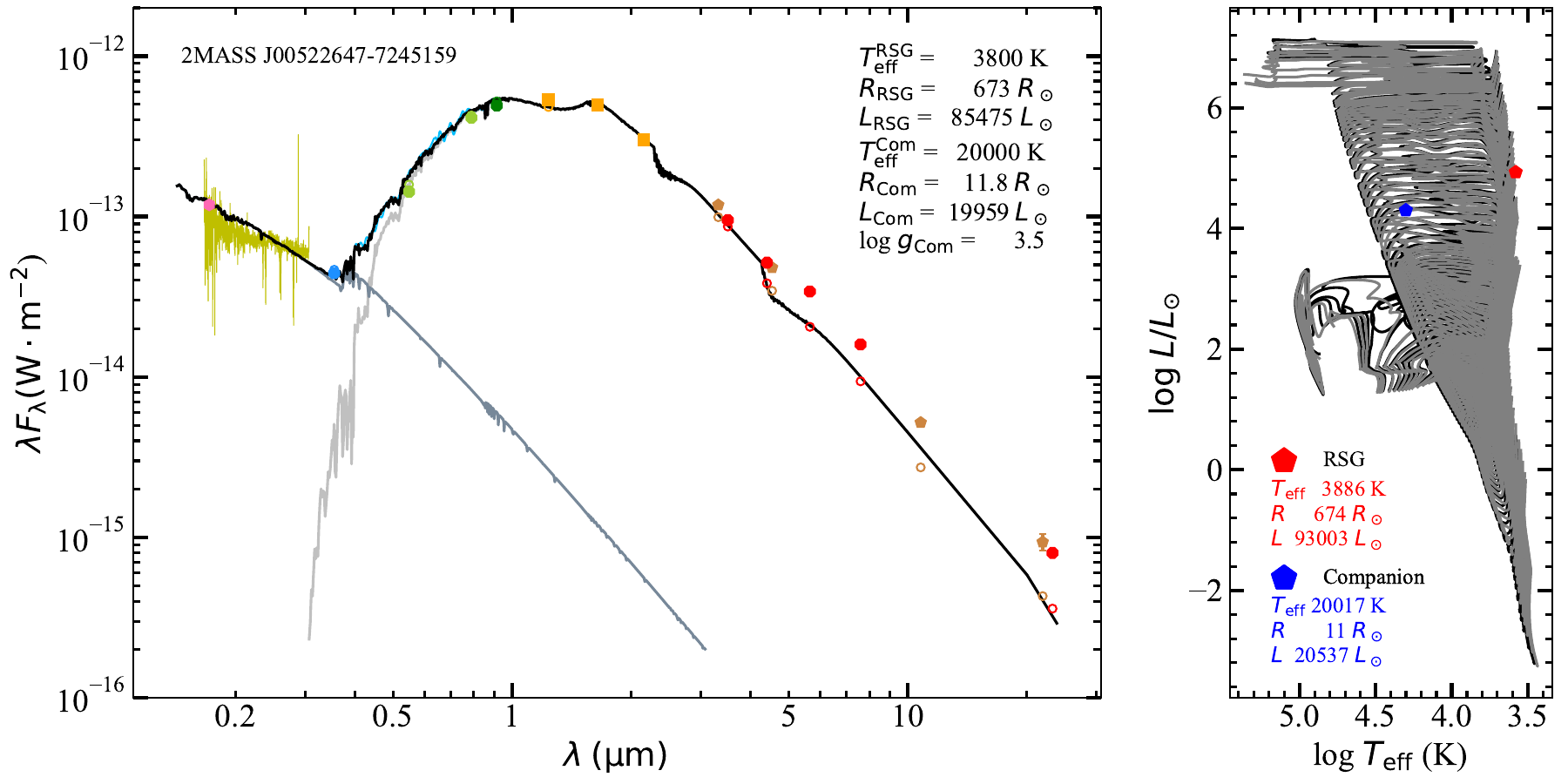}
	   \vspace{-0.0in}
	   \caption{\label{SED fitting}
	   An example of the SED fitting for the companion star. The left panel is similar to Figure \ref{RSG SED fitting}. The right panel shows the locations of RSG (red) and companion (blue) in the H-R diagram. The black and gray lines represent different metallicity of the PARSEC evolutionary tracks, at $Z$ = 0.004 and 0.006 for LMC, $Z$ = 0.001 and 0.002 for SMC \citep{2012MNRAS.427..127B}. }
	   \end{figure*}

\begin{figure*}
		  \centering
		  \vspace{-0.0in}
		  \includegraphics[angle=0,width=90mm]{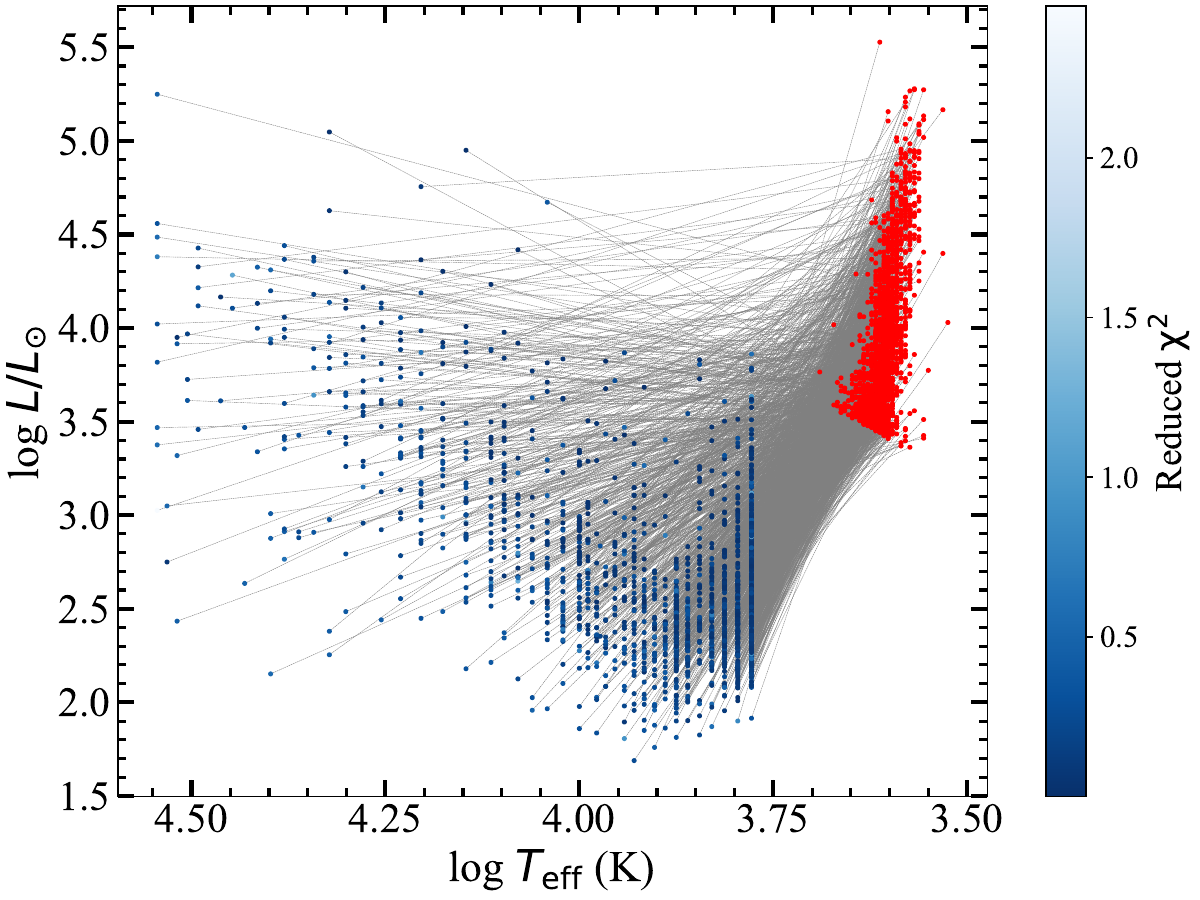}
		  \vspace{-0.0in}
		  \caption{\label{HR_of_companion} The H-R diagram of all the companion stars \textbf(blue dots) and their corresponding RSGs (red dots), gray connecting lines demonstrate the one-to-one correspondence between each stellar pair. The color bar represents the reduced $\mathrm{\chi^2}$ of the companion SED fitting.}
	   \end{figure*}

\begin{figure*}
	   	\centering
	   	\vspace{-0.0in}
	   	\includegraphics[angle=0,width=170mm]{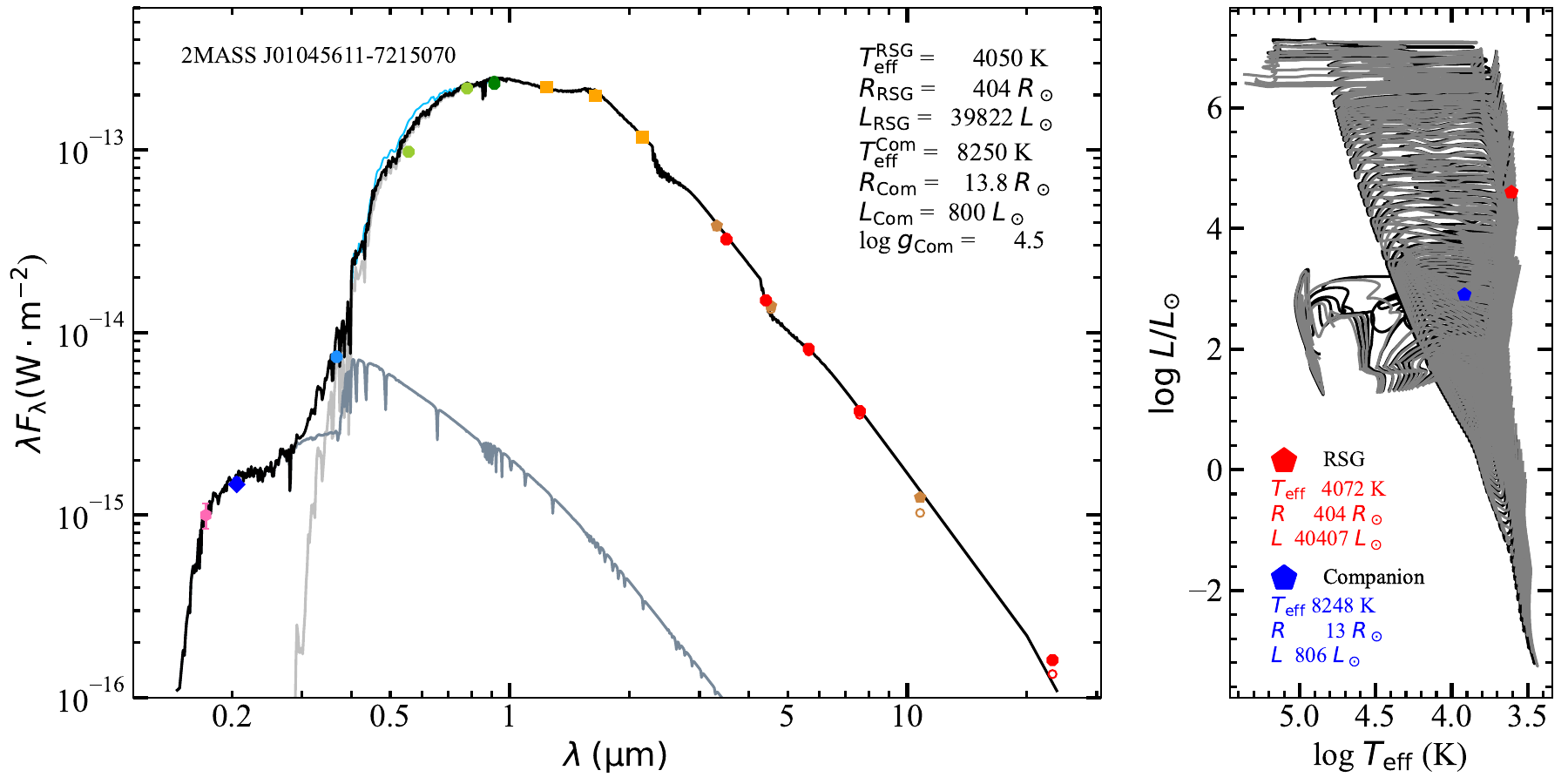}
	   	\vspace{-0.0in}
	   	\caption{\label{SED fitting_for_cool}
	   		It is similar to the Figure \ref{SED fitting}, but it is an example of the SED fitting for a "cool" with $T_\mathrm{eff}$ = 8250 K companion star.}
	   \end{figure*}

The HST/STIS spectra are searched from MAST to crosscheck the method. It is found that 15\footnote{2MASS J00513280-7205493, J00522647-7245159, J00532362-7247013, J00533967-7232089, J00534451-7233192, J00580864-7219270, J01010366-7202588, J01023794-7235547, J01024076-7217173, J01024208-7237294, J01031094-7218327, J01031855-7206461, J01052742-7217044, J01081478-7246411 and J01101248-7237310.} stars have UV spectra consistent with photometric data. As an example shown in Figure \ref{RSG SED fitting}, the yellow solid line represents the HST/STIS spectrum.  All 15 stars are also identified as binaries by the SED fitting method. The stellar parameters of the companions derived from the fitting with and without spectroscopic data are summarized in Table \ref{tab:HST_stellar_parameters}. The temperatures of companions derived from two methods are mostly in agreement, which means that our fitting for companions with photometric data is reliable. There are two objects with apparently different $T_\mathrm{eff}$, but their log $L/L_{\odot}$ are roughly the same. This can be understood that the stellar radius is coupled to stellar log $g$.

\subsection{Binary Fraction of RSGs with log $L/L_{\odot} > 4.0$} \label{subsec:binary_fraction_of RSGs_with_L>4.0}
The low-luminosity RSGs sample may be contaminated by asymptotic giant branch (AGB) stars \citep[e.g., see Figure 1 in ][]{2024ApJ...969...81Z}. In this regard, a purer subsample is obtained by constraining log $L/L_{\odot} > 4.0$, the same as in, e.g., \citet{2020ApJ...900..118N}. The subsample contains 1532 RSGs in the LMC and 666 RSGs in the SMC, and the binary fraction becomes 26.6\% $\pm$ 1.1\% (407/1532) and 26.4\% $\pm$ 1.7\% (176/666), respectively. This is 3.6\% and 5.8\% reduced in comparison with the total sample.

\subsection{Detectable Range of Binary Orbital Periods} \label{subsec:Range_Periods}
As previously mentioned, our methodology exhibits insensitivity to the orbital period of binary systems. However, we can effectively estimate the period within the detectable range. We adopted a methodology similar to that employed by \citet{2022MNRAS.513.5847P} for estimating the maximum orbital period of binary systems. Their approach utilizes cross-match distance (XMD), which can be regarded as analogous to aperture size, to determine the upper limits of binary orbital periods.	The median XMD of our multi-band photometric catalogues is $\sim$ 0.09$''$ - 0.4$''$(e.g., demonstrating median XMD values of 0.09$''$ between 2MASS and SMASH; 0.4$''$ between 2MASS and XMM-OM), yielding an upper limit of $\log P / [\text{d}]$ $\sim$ 8 for the binary orbital period, or $\sim$ 30,000 AU for the orbital separation. Concurrently, the intrinsic physical size constraints of RSGs establish a fundamental lower boundary for binary periods. Previous studies have indicated this minimum period threshold to be $\log P / [\text{d}]$ $\sim$ 2.3 \citep{2020AA...635A..29P}. Consequently, the binary orbital period range derived through our methodology spans approximately 2.3 < $\log P / [\text{d}]$ < 8.

The lower binary fraction of RSGs ($\sim$ 30\%) we obtained compared to their progenitors ($\approx$ 50\% - $\approx$ 70\% for OB-type stars) can be reasonably explained through orbital period evolution. The higher binary fraction of OB-type stars is determined through multi-epoch spectroscopy surveys sensitive to orbital periods of  $0 \lesssim \log P / [\text{d}]  \lesssim 3$, whereas our detection threshold for RSGs corresponds to a minimum orbital period of $\log P$ $\sim$ 2.3 days. Systems with shorter orbital periods than this threshold may have undergone binary merger events during their evolutionary history, effectively converting potential binary systems into single stars. In addition, our derived binary fraction for RSGs generally exceeds those obtained through RV method. This discrepancy arises because our methodology probes a wider range of orbital periods compared to the more limited period sensitivity range of conventional RV techniques (see Section \ref{sec:intro}, $3.3 \leq \log P / [\text{d}]  \leq 4.3$). The extended period coverage in our study enables detection of binary systems that would remain undetected through RV alone. In fact, the fraction of triples among massive stars is considerable \citep{2017ApJS..230...15M}. The third body significantly influences the evolution of the inner binary \citep{2020A&A...640A..16T,2023A&A...678A..60K}. Combined with a binary fraction of $\sim$ $50 - 70\%$ within the period range of $0 \lesssim \log P / [\text{d}]  \lesssim 3$, our study suggests that $\sim$ $1/3$ massive stars hosts a third companion within $\sim$ 30,000 AU. In addition, the series of studies by Neugent et al. in section \ref{sec:intro} did not explicitly specify their detectable period range. Considering a typical photometric and spectroscopic apertures (e.g. 0.5$''$ ), their detectable period coverage could be similar to ours.

\subsection{Comparison with Previous Work on the Fraction of Binary} \label{subsec:Comparison}

We compare the binary fractions of RSGs in the LMC \citep{2020ApJ...900..118N} and the SMC \citep{2022MNRAS.513.5847P}.

\begin{figure*}
	   	\centering
	   	\vspace{-0.0in}
	   	\includegraphics[angle=0,width=170mm]{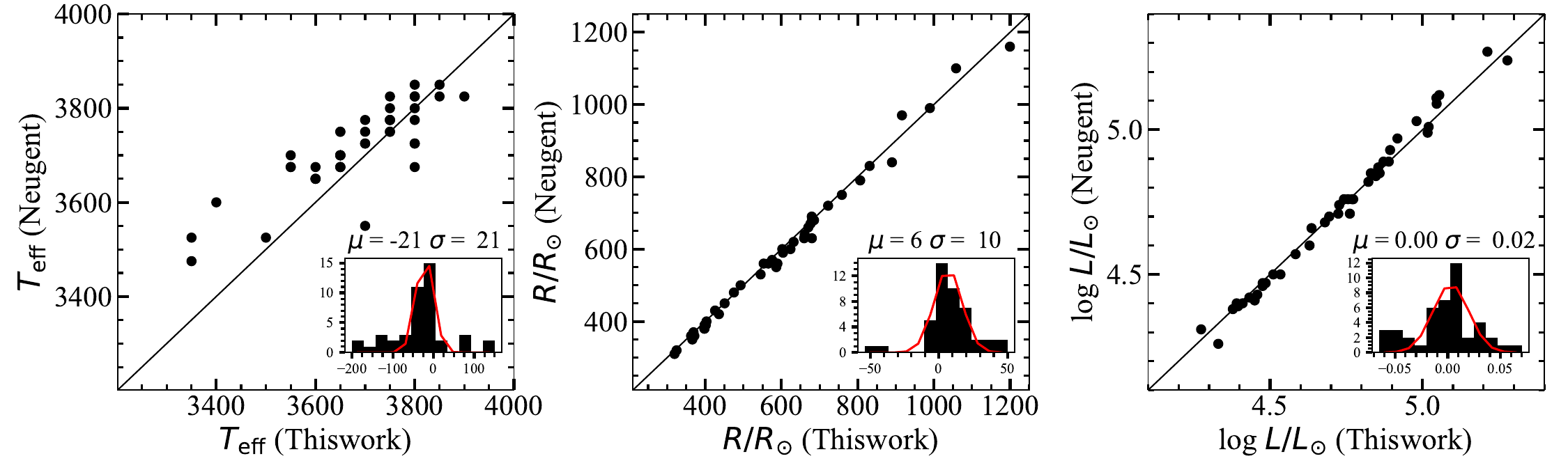}
	   	\vspace{-0.0in}
	   	\caption{Comparison of $T_{\mathrm{eff}}$, $R$ and $L$ derived from this work and \citet{2020ApJ...900..118N}. \label{Comparison_This_Neugent_APOOGE}}
	   \end{figure*}

\citet{2020ApJ...900..118N} use a $k$-nearest neighbors algorithm to calculate the binary fraction of RSG with log$L/L_{\odot} > 4$ for a sample of 1752 RSGs with likelihood of binarity in the LMC. The binary fraction of RSGs with O- or B-type companions is $13.5 ^{+7.56} _{-6.67}\%$ \citep{2020ApJ...900..118N}. By cross-matching with their sample, 1481 stars are common. For these 1481 stars, the binary fraction of \citet{2020ApJ...900..118N} is 11.55\% assuming that stars with likelihood $\geq$ 50\% are binaries, meanwhile the binary fraction of this work is 26.47\%, much higher than theirs. Of the 171 stars with likelihood greater than 50\%, 133 stars (78\%) are also identified by this work. It is worth noting that there are 6 stars\footnote{2MASS J04543854-6911170, J04595731-6748133, J05170897-6932211, J05284548-6858022, J05292757-6908502 and J05414402-6912027.} which are binaries with 100\% probability by \citet{2020ApJ...900..118N}, but are classified as single stars in this work. However, this does not affect the overall fraction, as the specific discrepancies observed in these 6 stars are detailed in Appendix \ref{technical_details}. In addition, a comparison of the stellar parameters ($T_\mathrm{eff}$, $R$ and $L$) of 44 common RSGs is shown in Figure \ref{Comparison_This_Neugent_APOOGE}. The $T_\mathrm{eff}$, $R$ and $L$ determined spectroscopically by \citet{2020ApJ...900..118N} are in general agreement with those from this work, with the $\mu$ and $\sigma$ of residuals shown by insets in Figure \ref{Comparison_This_Neugent_APOOGE}.

\citet{2022MNRAS.513.5847P} report a binary fraction of $15.7\pm 1.5\%$ for RSGs in the SMC by the detection in the UVIT/F172M band because single RSGs are non-detectable at this wavelength. In their sample of 862 RSG candidates in the SMC, 560 RSGs are located in the footprint of the UVIT survey. They find 88 stars detected in the F172M band and attribute to the presence of hot companions. By cross-matching with the 862 RSG candidates of \citet{2022MNRAS.513.5847P}, 654 RSGs are common. The binary fraction of this sample is 28.75\% in this work, which is about twice of 15.7\%. This can be understood because the F172M filter is sensitive only to very hot stars as the effective wavelength corresponds to a Wien temperature of $\sim$ 17,000K. For the 88 stars detected in UVIT/F172M, 84 of them are included in this work. Except one has the reduced $\chi^2$ above the 95\% confidence level, the other 83 stars are all binaries identified by this work. This implies that our method can identify more than 95\% binaries with UV excess. It also means that a small fraction is missed during our removal of poor SED fitting caused by the UV excess. 

\subsection{UV Excess in the Color-color Diagrams} \label{subsec:Color-color_diagrams}

\begin{figure*}
	   \centering
	   \vspace{-0.0in}
	   \includegraphics[angle=0,width=170mm]{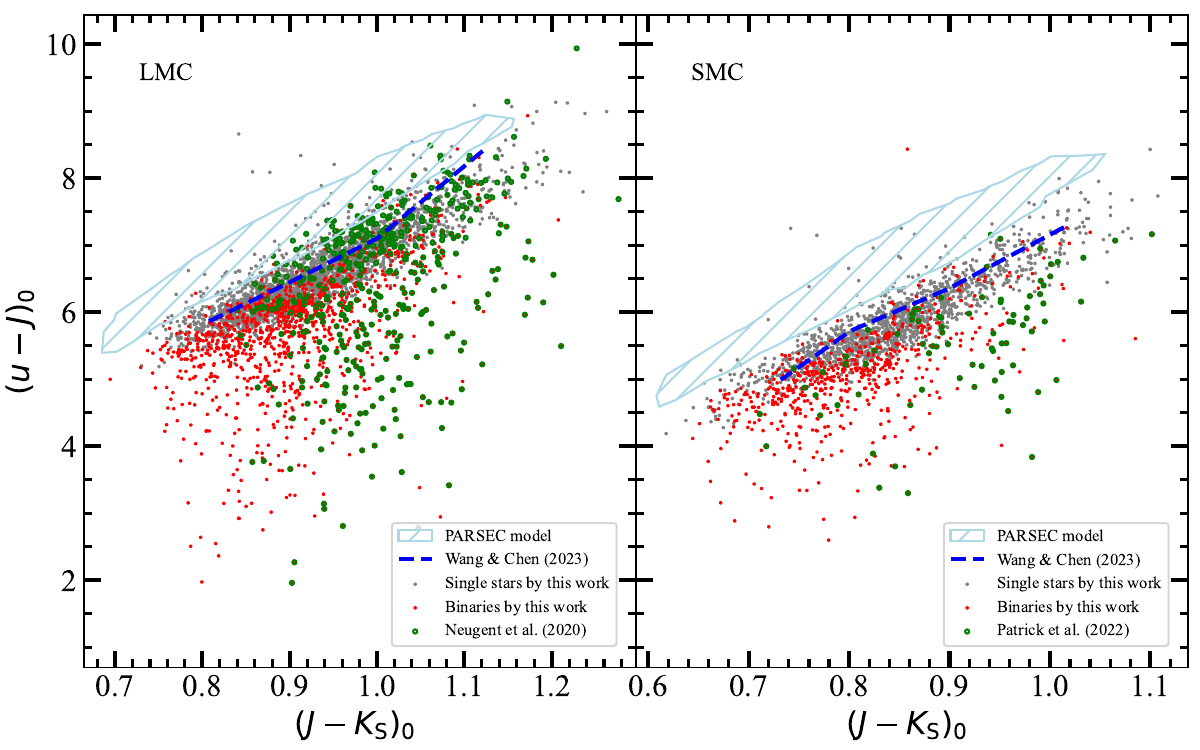}
	   \vspace{-0.0in}
	   \caption{The RSGs sample of this work in the $(u - J)_{0}$ vs. $(J - K_{\mathrm{S}})_{0}$ diagram. The red dots represent the RSG binaries identified by this work, and the green circles represent the RSG binaries identified by \citet{2020ApJ...900..118N} and \citet{2022MNRAS.513.5847P}. The gray dots represent the RSG single stars in this work. The region of light blue shaded area represents the location of single RSG stars  predicted by the PARSEC model, and the blue dashed line represents the location of RSGs by \citet{2023ApJ...946...43W}.
	   \label{ccd_smash_2mass}}
	   \end{figure*}

Theoretically, RSG with hot companion should have smaller $(\mathrm{UV} - \mathrm{NIR})_{0}$ than single RSG because of the higher UV emission by the companion. Then, the RSG binaries should appear below the main single branch in the $(\mathrm{UV} - \mathrm{NIR})_{0}$ vs. $(J - K_{\mathrm{S}})_{0}$ diagram since the NIR color is determined only by RSG. Figure \ref{ccd_smash_2mass} shows the distribution in the $(u - J)_{0}$ vs. $(J - K_{\mathrm{S}})_{0}$ diagram for the RSG stars in the LMC (left) and SMC (right). The red solid dots and the gray circles denote the RSG binaries and singles respectively identified in this work. Clearly, there is a concentrated narrow band at the upper side, which is consistent with two previous results displayed, i.e. the light-blue shaded area for the RSG single stars calculated by the PARSEC model \citep{2012MNRAS.427..127B}, and the blue dashed line for the RSG sample of \citet{2023ApJ...946...43W}. It can be seen that there is a systematic shift between them, with the result of \citet{2023ApJ...946...43W} closer to the main population of the sample in this work. The reason for the difference may be the presence of chromospheric activity of RSGs, which leads to smaller UV fluxes predicted by the PARSEC model \citep{2019IAUS..343..365C}. Meanwhile, there are some binaries in the single star sequence. Though these binaries have a hotter companion, the companion is not so hot to augment the flux in the SMASH $u$ band significantly. In addition, there are a few stars beyond the upper bound of the PARSEC model, which may be caused by incorrect photometry in the SMASH $u$ band.

\begin{figure*}
		  \centering
		  \vspace{-0.0in}
		  \includegraphics[angle=0,width=170mm]{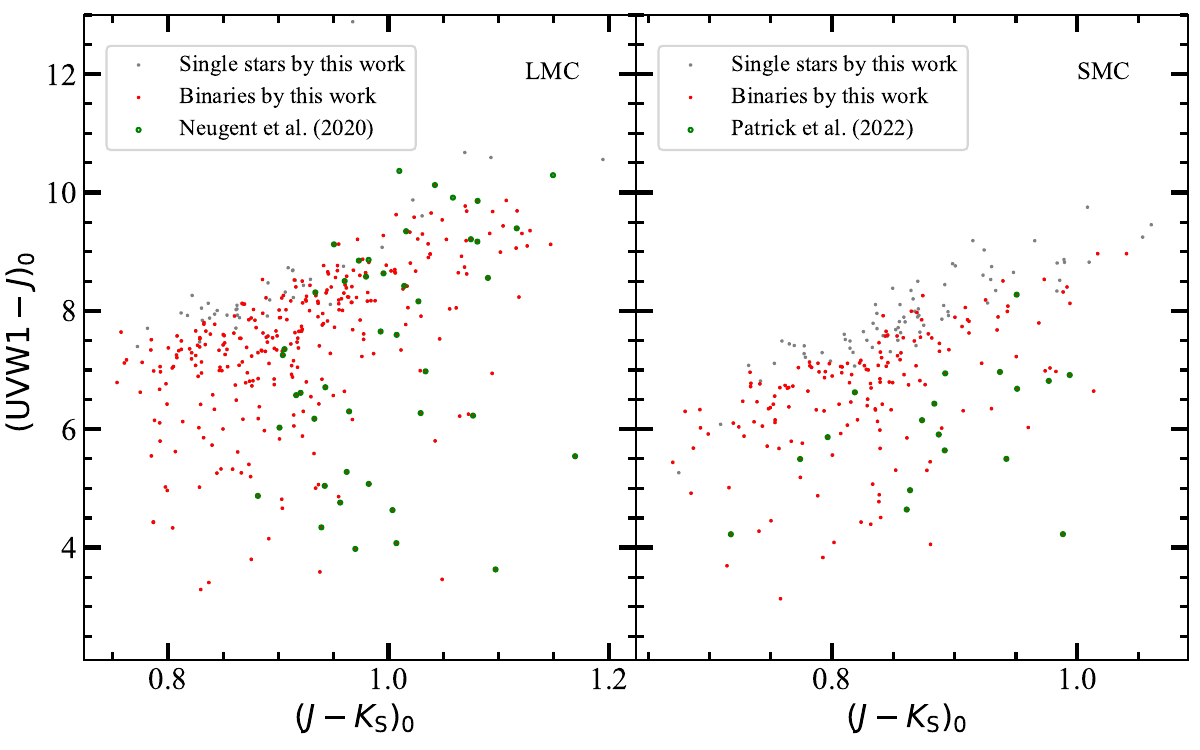}
		  \vspace{-0.0in}
		  \caption{The RSG binaries from the sample of this work in $(\mathrm{UVW1} - J)_{0}$ vs. $(J - K_{\mathrm{S}})_{0}$ diagram. This figure is similar to Figure \ref{ccd_smash_2mass}. 
		  \label{2ccd_xmm_2mass}}
	   \end{figure*}

As expected, the stars below the main branch are mostly binaries. For the stars detected in the SMASH $u$ band, the binary fraction is 31.57\% in the LMC and 32.54\% in the SMC. By replacing the SMASH $u$ band by the UVW1 band, a similar color-color diagram is presented in Figure \ref{2ccd_xmm_2mass} as the $(\mathrm{UVW1} - J)_{0}$ vs. $(J - K_{\mathrm{S}})_{0}$ diagram. Then, the binary fraction is up to 88.38\% for the LMC and 70.8\% for the SMC in the RSG sample detected in the UVW1 band. The great increase in the UVW1 band is understandable since the effective wavelength is shorter for the UVW1 with $\lambda_{\rm eff}\sim0.29 \mu$m band than the SMASH $u$ band with $\lambda_{\rm eff}\sim0.38 \mu$m. The small shift to UV band greatly increases the detection efficiency of hot companions for RSGs, which may guide to the future observational preference.

\subsection{Additional Evidences and Limitations} \label{subsec:Additional_evidence_Limitations}

In addition to the SED fitting method, the RSG binaries are also searched by the light curve (LC) and RV methods. It is found that 13 LCs have significant eclipsing characteristic after cross-matching with the eclipsing binary catalog of \citet{2007AJ....134.1963F} and \citet{2016AcA....66..421P}. These stars are all identified as binaries in this work.

For the RV method, the APOGEE provides an opportunity for its high spectral resolution and multiple visits to one object. The epoch RV data are taken from APOGEE/DR17. The method is similar to previous studies to calculate the RV variation \citep{2013A&A...550A.107S,2015A&A...580A..93D,2019AA...624A.129P,2021MNRAS.502.4890D}. The stars with significant variation are firstly selected by:
\begin{equation}
	\frac{|v_i-v_j|}{\sqrt{\sigma_i^2+\sigma_j^2}}>3 \text{,}
	\label{eq:significance}
\end{equation}
where $v_i$ and $v_j$ are any two epochs RVs of the source, and $\sigma_i$ and $\sigma_j$ are the corresponding uncertainties. Moreover, the $|\Delta V_{\mathrm{max}}|$ is defined as the maximum difference of RV, and the binary criterion of $|\Delta V_{\mathrm{max}}| >$ 11 km s$^{-1}$ is adopted as suggested by \citet{2021MNRAS.502.4890D}. This yields 10 RSG binaries in our sample. Among the 10 RSG binaries, two are identified by the UV excess in this work. In addition, Gaia DR3 provides $|\Delta V_{\mathrm{max}}|$ of RV as well, but no epoch RV. Due to lack of epoch RV data, only the criterion of $|\Delta V_{\mathrm{max}}| >$ 11 km s$^{-1}$ is adopted, which results in 355 binary candidates. Among the 355 binary candidates, 92 are identified by the UV excess in this work, a much larger number than the APOGEE result. The reason may come from two sides. One is that the Gaia criterion is not so strict so that the candidates are contaminated significantly by single stars. The other is that our estimation can only find the binaries with hot companions while the RV method is not limited by this factor. The binaries identified by the LC and RV methods are summarized in Table \ref{tab:ex_evidence}.

In this work, our method relies on the UV observations, depending on detection of the excess UV fluxes to identify the presence of hot companion. The RSG binaries with cool companions cannot be identified in this way because the fluxes of cool companion will be mixed with RSGs as evidenced by the lower limit of $T_\mathrm{eff} =$ 6000 K of the companion. In addition, some stars may be absent of UV photometry or the UV fluxes of companion may be eclipsed by RSG. Therefore, our results should be taken as the lower limit of RSG binary fraction.

\subsection{Potential Contamination of Binary Fraction} \label{subsec:Contamination_Binary_fraction}
\subsubsection{UV excess by Circumstellar Dust}
UV excess can be caused by sufficient circumstellar dust of RSG, which is indicated by the linear relation of the NUV excess with dust production rate $\dot{M}_{\rm{d}}$ and extra extinction \citep{2005ApJ...634.1286M}. The average $\dot{M}_{\rm{d}}$ are on the order of $10^{-10}$ $\rm{M}_\odot$$yr^{-1}$ and $10^{-11}$ $\rm{M}_\odot$$yr^{-1}$ in the LMC and SMC, respectively \citep{2012ApJ...753...71R,2012ApJ...748...40B,2016MNRAS.457.2814S,2024AJ....167...51W}. With such low dust production rate due to the low metallicity of MCs, the NUV excess from circumstellar dust scattering is expected to be negligible. Thus the circumstellar dust scattering can not account for the observed UV excess.

\begin{figure*}
		\centering
		\vspace{-0.0in}
		 \includegraphics[angle=0,width=85mm]{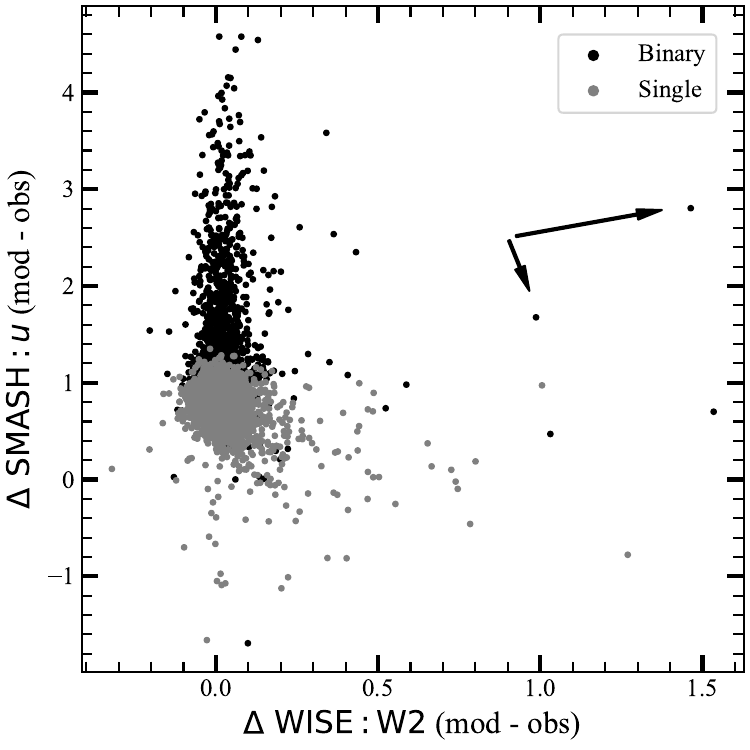}
		\vspace{-0.0in}
		\caption{\label{UV excess by dust} IR excess vs. UV excess diagram. The two stars indicated by black arrows may be affected by circumstellar dust.}
	 \end{figure*}

In addition, the UV excess if caused by circumstellar dust should correlate with the IR excess. However, as shown in Figure \ref{UV excess by dust}, there is no correlation between the UV excess represented by the modeled magnitude minus the observed for SMASH $u$ and the IR excess represented by the modeled magnitude minus observed for WISE W2. The distribution of IR excess is concentrated for the most stars, except two stars\footnote{2MASS J05482065-7346202 and 04490798-6907191.} pointed by the black arrows. Therefore, the effect of circumstellar dust on binary fraction is small in this work.

\subsubsection{Contamination Probability from Line of Sight}
As mentioned in the work of \citet{2020ApJ...900..118N} and \citet{2021ApJ...908...87N}, a possible contaminant is line of sight pairings. This means that the UV excess of RSGs do not come from real hot companion, but from foreground or background stars of the MCs. They find a contamination rate of $1.9\%\pm{2.0\%}$ for OB companions. To calculate the contamination rate in this work, a similar method is applied as described in \citet{2020ApJ...900..118N}. Firstly, the OBA stars in the studied area are selected from Gaia DR3 by the intrinsic color
$(G_{\rm{BP}} - G_{\rm{RP}})_{0}$ and $T_\mathrm{eff}$. We set $T_\mathrm{eff}$ = 7400 K as the lower limit of A type stars, and the corresponding $(G_{\rm{BP}} - G_{\rm{RP}})_{0}$ is calculated as = 0.44 and 0.38 in the LMC and SMC, respectively \citep{2020ApJ...891..137Z,2023ApJ...945..132C}. An average $E(B - V)$ = 0.13 and 0.09 are adopted \citep{2007AJ....133.2393M} and converted to $E(G_{\rm{BP}} - G_{\rm{RP}})$ with the extinction law by \citet{2019ApJ...877..116W,2023ApJ...946...43W}. Then, for each RSG binary determined in Section \ref{subsec:Identification_Binary}, the locations of all OBA stars are searched within a 5$'$ radius of the RSG binary. Finally, we randomly move the binary to another position within 5$'$. If the new position is within 1$''$ of any OBA stars, it means a line of sight pairing. An MCMC code\footnote{\url{https://github.com/KNeugent/LineOfSightBinaries}} is run 10,000 times for each RSG binary. A $4.33\%\pm{1.84\%}$ contamination rate is found in this work. This means that the contamination rate is $4.33\%\pm{1.84\%}$ for the whole sample, while $1.9\%\pm{2.0\%}$ for the sample with log $L/L_{\odot} > 4.0$.

\section{summary} \label{subsec:summary}

In this work, we calculate the binary fraction of RSGs in the LMC and SMC by the SED fitting method in the sample of 4695 (LMC) and 2097 (SMC) RSGs. After fitting the SEDs of RSGs, the RSG binaries are identified by comparing the observed flux with the modeled flux of the RSGs in the UV bands. It is found that the binary fraction of RSGs is 30.2\% $\pm$ 0.7\% in the LMC and 32.2\% $\pm$ 1\% in the SMC for whe whole sample, while it is 26.6\% $\pm$ 1.1\% in the LMC and 26.4\% $\pm$ 1.7\% in the SMC for sample of RSGs with log $L/L_{\odot} > 4.0$. The detectable period range associated with our methodology spans $\sim$ 2.3 < $\log P / [\text{d}]$ < $\sim$ 8. In addition, the possible contamination rate by chance sight line pairing is $4.33\%\pm{1.84\%}$ for the whole sample and $1.9\%\pm{2.0\%}$ for the sample with log $L/L_{\odot} > 4.0$. Due to the limitations of method and data, this result should be the lower limit of binary fraction in the RSGs. In the process of the SED fitting, the stellar parameters, i.e. the $T_{\mathrm{eff}}$, $R$ and $L$ of the RSGs, and the $T_{\mathrm{eff}}$, $R$, $L$ and log $g$ of the companions are derived. Among the binary RSGs identified by the SED fitting, 15 are supported by fitting the HST/STIS spectra. In addition to the SED fitting method, the LC and RV methods are used to confirm our identification of binaries for a small part of the sample.

\section*{Acknowledgements}
We are grateful to the anonymous referee for his very helpful comments and suggestions to improve this paper. We are also grateful to Drs. Xiaodian Chen, Jun Li, Yi Ren and Ming Yang for their helpful discussion and suggestion. This work is supported by the NSFC project 12133002 and 12373028, National Key R\&D Program of China No. 2019YFA0405500, and CMS-CSST-2021-A09. S.W. acknowledges the support from the Youth Innovation Promotion Association of the CAS (grant No. 2023065). The numerical computations were conducted on the Qilu Normal University High Performance Computing, Jinan Key Laboratory of Astronomical Data. This work has made use of the data from SMASH, 2MASS, Gaia, XMM-OM, $Swift$/UVOT, UVIT, HST, GALEX, SMSS, SPITZER, WISE surveys.

\section*{Data Availability}

The data underlying this article are available in the article and in its online supplementary material. The data underlying this article will be shared on reasonable request to the corresponding author.



\bibliographystyle{mnras}
\bibliography{binary_lmc_smc} 


	\begin{table*}
	\caption{The information of all filters used for constructing SED.} 
	\label{passbandinfo} 
	\begin{tabular}{cc}
	   \hline \hline
	   Facility/Instrument & Filter(effective wavelength, unit in $\mu m$) \\
	   \hline
	   HST & F170W(0.19), F225W(0.24), F275W(0.27), F280N(0.28), F300W(0.30), F336W(0.33), F343N(0.34), F373N(0.37)\\
	   GALEX & FUV(0.15), NUV(0.23)\\
	   UVIT &  F154W(0.15), F169M(0.16), F172M(0.17), N219M(0.22), N245M(0.24), N263M(0.26), N279N(0.28) \\
	   $Swift$/UVOT & UVW2(0.21), UVM2(0.22), UVW1(0.28), $U$(0.35), $B$(0.44), $V$(0.55)\\
	   XMM-OM & UVW2(0.20), UVM2(0.23), UVW1(0.29), $U$(0.35), $B$(0.43), $V$(0.54) \\
	   SMASH & $u$(0.36), $g$(0.47) \\
	   SMSS  & $u$(0.35), $v$(0.39), $g$(0.50), $r$(0.61), $i$(0.77), $z$(0.91) \\
	   APASS & Johnson: $B$(0.43), $V$(0.54); SDSS: $g’$(0.46), $r’$(0.61), $i’$(0.74) \\
	   Gaia  & $G_{\rm BP}$(0.50), $G_{\rm RP}$(0.76) \\
	   2MASS & $J$(1.24), $H$(1.66), $K_{\rm S}$(2.16) \\
	   SPITZER & IRAC: [3.6] (3.51), [4.5] (4.44), [5.8] (5.63), [8.0] (7.59); MIPS: [24] (23.21) \\
	   WISE    & W1(3.35), W2(4.60), W3(11.56), W4(22.09)\\
	   \hline
	   \multicolumn{2}{l}{\footnotesize{$^a$ The effective wavelength for Vega is taken from SVO Filter Profile Service\footnote{\url{http://svo2.cab.inta-csic.es/svo/theory/fps3/index.php?mode=browse}}.}}\\

	\end{tabular}
	\end{table*}
	   
\begin{table*}
\caption{Stellar parameters of RSGs in the Magellanic Clouds. A full version of this table is available as supplementary material.}
\label{tab:Parameters of RSGs and companions}
\begin{tabular}{ccccccccc}
\hline\hline    
2MASS  & \multicolumn{3}{c}{RSG} &\multicolumn{4}{c}{Companion} & Type$^{d}$ \\
   &  $T_\mathrm{eff}$$^{a}$ &  $R/R_{\odot}$$^{b}$  & log $L/L_{\odot}$$^{c}$  & $T_\mathrm{eff}$ &  $R/R_{\odot}$  & log $L/L_{\odot}$   & log $g$  &  \\ 
   \multicolumn{9}{c}{LMC} \\
\hline
   04160129-7136083 & 3750 & 289 & 4.17 &  &  &  &  & S \\
   04191178-6544192 & 3800 & 126 & 3.47 &  &  &  &  & S \\
   ... & ... & ... & ... & ... & ... & ... & ... & ... \\
   04182834-7741567 & 4150 & 143 & 3.74 & 7500 & 11 & 2.53 & 0.5 & B \\
04262486-6618046 & 4250 & 122 & 3.64 & 8500 & 10 & 2.64 & 4.5 & B \\
\hline
\multicolumn{9}{c}{SMC} \\
\hline
00082691-7216537 & 4200 & 304 & 4.42 &  &  &  &  & S \\
   00105622-7520144 & 4250 & 108 & 3.54 &  &  &  &  & S \\
   ... & ... & ... & ... & ... & ... & ... & ... & ... \\
01411596-7333447 & 4200 & 139 & 3.74 & 10500 & 8 & 2.86 & 2.0 & B \\
01421678-7316371 & 3700 & 431 & 4.5 & 6000 & 29 & 2.98 & 2.5 & B \\
\hline
\multicolumn{9}{l}{\footnotesize{$^a$ Typical uncertainty 50 K.}}\\
\multicolumn{9}{l}{\footnotesize{$^b$ Typical relative uncertainty $^{+1.4} _{-2.2} \%$.}}\\
\multicolumn{9}{l}{\footnotesize{$^c$ Typical relative uncertainty $^{+1.36}_{-1.37}\%$.}}\\
\multicolumn{9}{l}{\footnotesize{$^d$ S = single star, B = binary system.}} \\
\multicolumn{9}{l}{\footnotesize{The unit of $T_\mathrm{eff}$ is K, $R$ is $R_{\odot}$ and $L$ is $L_{\odot}$.}}\\
\end{tabular}
\end{table*}
	   
\begin{table*}
\caption{Binary fraction of RSGs.}
\label{tab:Reported binary fraction of RSGs}
\begin{tabular}{cccc}
\hline\hline
Region & Fraction & Method & Reference\\

\textit{LMC} &	\textbf{30.2$\%$ / 26.6$\%$}$^{a}$ &  \textit{SED fitting} & \textit{This work} \\
LMC &$ 19.5 ^{+7.6}_{-6.7}$ $\%$ & Machine learning based on spectrum& \protect\cite{2020ApJ...900..118N} \\
LMC 30 Doradus   & Up limit 30$\%$ & RV & \protect\cite{2019AA...624A.129P} \\
 LMC  &14 $\pm$ 5 $\%$ & RV & \cite{2021MNRAS.502.4890D} \\
\hline
\textit{SMC} &\textbf{32.2$\%$ / 26.4$\%$}$^{b}$ &  \textit{SED fitting} & \textit{This work} \\
SMC&15.7$\pm$ 1.5 $\%$ & Detected FUV flux & \protect\cite{2022MNRAS.513.5847P} \\
SMC NGC 330  &30 $\pm$ 10 $\%$ &  RV & \protect\cite{2020AA...635A..29P} \\
 SMC   &15 $\pm$ 4 $\%$ &  RV & \cite{2021MNRAS.502.4890D} \\
 \hline
M31& Internal $41.2^{+12.0}_{-7.3}$ External $15.9^{+12.4}_{-1.9}$ &Machine learning based on spectrum & \protect\cite{2021ApJ...908...87N} \\
M33& $33.5^{+8.6}_{-5.0} \%$ & Machine learning based on spectrum & \protect\cite{2021ApJ...908...87N}\\
\hline                                   
\multicolumn{4}{l}{\footnotesize{$^a$ The former is the fraction for the whole sample, and the latter is for the subsample with log $L/L_{\odot} > 4.0$.}}\\
\multicolumn{4}{l}{\footnotesize{$^b$ The same as `a'.}}\\
\end{tabular}
\end{table*}
	   
	\begin{table*}
	\caption{Stellar parameters of the companions derived from photometry and spectrum.}
	\label{tab:HST_stellar_parameters}
	\begin{tabular}{ccccccc}    
	\hline\hline
	2MASS  & \multicolumn{2}{c}{$T_\mathrm{eff}$} &  \multicolumn{2}{c}{$R/R_{\odot}$}  & \multicolumn{2}{c}{log $L/L_{\odot}$}\\
	&by photometry&by spectrum&by photometry&by spectrum&by photometry&by spectrum\\
	\hline
	   00513280-7205493 & 16000 & 18000 & 14 & 12 & 4.08 & 4.13 \\
	   00522647-7245159 & 17000 & 20000 & 14 & 12 & 4.2 & 4.3 \\
	   00532362-7247013 & 17000 & 21000 & 9 & 6 & 3.81 & 3.84 \\
	   00533967-7232089 & 14000 & 15000 & 13 & 12 & 3.75 & 3.81 \\
	   00534451-7233192 & 15000 & 15000 & 13 & 13 & 3.89 & 3.9 \\
	   00580864-7219270 & 14000 & 13000 & 4 & 5 & 2.8 & 2.76 \\
	   01010366-7202588 & 9500 & 16000 & 19 & 6 & 3.42 & 3.33 \\
	   01023794-7235547 & 15000 & 19000 & 13 & 9 & 3.9 & 4.0 \\
	   01024076-7217173 & 14000 & 19000 & 12 & 7 & 3.68 & 3.72 \\
	   01024208-7237294 & 10500 & 12500 & 9 & 7 & 2.9 & 3.06 \\
	   01031094-7218327 & 11500 & 15000 & 14 & 8 & 3.48 & 3.52 \\
	   01031855-7206461 & 11000 & 17000 & 19 & 11 & 3.68 & 3.98 \\
	   01052742-7217044 & 19000 & 21000 & 8 & 7 & 3.91 & 3.92 \\
	   01081478-7246411 & 10500 & 8250 & 8 & 21 & 2.79 & 3.27 \\
	   01101248-7237310 & 14000 & 14000 & 6 & 7 & 3.13 & 3.17 \\
	\hline
	\multicolumn{7}{l}{\footnotesize{The unit of $T_\mathrm{eff}$ is K, $R$ is $R_{\odot}$ and $L$ is $L_{\odot}$.}}\\
	\end{tabular}
	\end{table*}
	   
	\begin{table*}
	\caption{RSG binaries identified by the LC and RV methods. A full version of this table is available as supplementary material.}
	\label{tab:ex_evidence}
	\begin{tabular}{cccccc}
	   \hline\hline
	   
	   2MASS & method$^{a,\ b}$ & $|\Delta V_{\mathrm{max}}|^{c}$ & $|\Delta V_{\mathrm{max}}|^{d}$ & This work$^{e}$\\
	   \hline
	   \multicolumn{5}{c}{LMC} \\
	   \hline
		  04444215-6924427 & LC[a] &  &  & B \\
		  04455518-6915438 & LC[a] &  &  & B \\
		  ... & ... & ...  & ... & ... \\
		  04355829-7142264 & RV & 16.7 &  &  \\
		  06000792-6809109 & RV &  & 13.1 &S \\
	   \hline
	   \multicolumn{5}{c}{SMC} \\
	   \hline
		  01000495-7246106 & LC[a] &  &  & B \\
		  01002813-7303072 & LC[a] &  &  & B \\
		  ... & ... & ...  & ... & ... \\
		  01303395-7318417 & RV  &  &16.7 &S \\
		  01330779-7402383 & RV &  & 12.0 & S\\
	   \hline
	   \multicolumn{6}{l}{\footnotesize{$^a$ From \citet{2016AcA....66..421P}.}}\\
	   \multicolumn{6}{l}{\footnotesize{$^b$ From \citet{2007AJ....134.1963F}.}}\\
	   \multicolumn{6}{l}{\footnotesize{$^c$ $|\Delta V_{\mathrm{max}}|$ from APOGEE DR17.}}\\
	   \multicolumn{6}{l}{\footnotesize{$^d$ $|\Delta V_{\mathrm{max}}|$ from Gaia DR3.}}\\
	   \multicolumn{6}{l}{\footnotesize{$^e$ S = single star, B = binary system, null value = star with $\chi^{2}$ above the 95\% confidence level.}}\\
	\end{tabular}
	\end{table*}
	   
\appendix
\section{Stellar Classification Comparison of Six Stars}
\label{technical_details}
\begin{figure*}
	\centering
	\vspace{-0.0in}
	\includegraphics[angle=0,width=170mm]{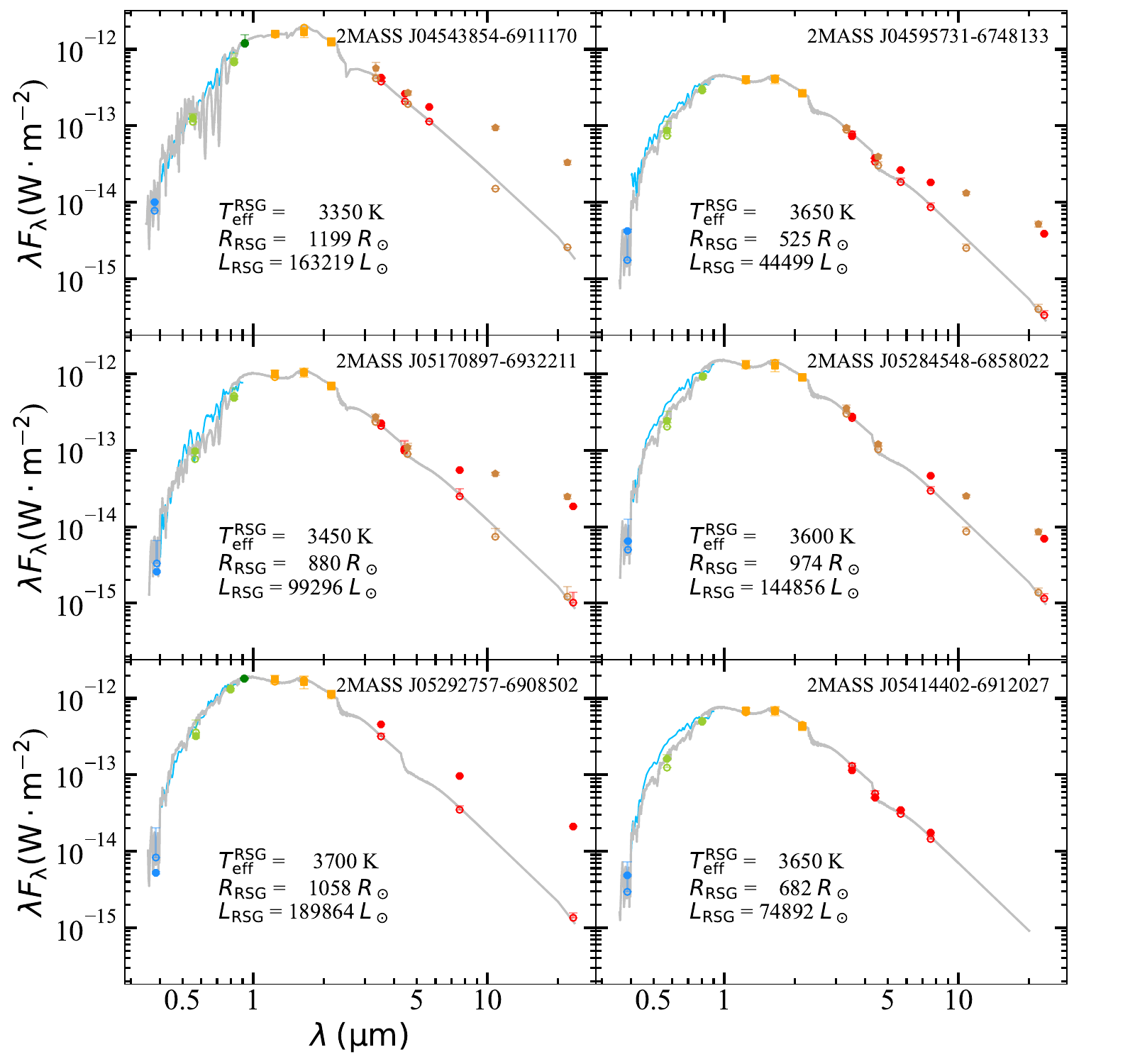}
	\vspace{-0.0in}
	\caption{The SED fitting of the RSG component of 2MASS J04543854-6911170, J04595731-6748133, J05170897-6932211, J05284548-6858022, J05292757-6908502 and J05414402-6912027. The symbol conventions follow Figure \ref{RSG SED fitting}.\label{nuegent_five}}
\end{figure*}

Of these, 2MASS J04543854-6911170 and J05292757-6908502 are identified as binaries by direct spectroscopy by \citet{2020ApJ...900..118N}, and should be reliable binary. In such rare conflict, it may be due to that the photometric epoch is not at the right orbital phase to expose the UV excess. For the other four stars, the Magellanic Clouds Photometric Survey \citep[MCPS,][]{2002AJ....123..855Z,2004AJ....128.1606Z} $U$ band photometric data used in \citet{2020ApJ...900..118N} is discrepant with the SMASH $u$ band data used in this work. In detail, the difference between the two observations, SMASH $u$ $-$ MCPS $U$, is 3.49, 2.76, 2.37 and 3.29 mag for 2MASS J04595731-6748133, J05170897-6932211, J05284548-6858022 and J05414402-6912027, while the systematic difference between the two bands is $\mu$ = 0.81 with $\sigma$ = 0.16 due to that MCPS uses the Vega system and SMASH uses the AB system. Because MCPS uses a 1-m telescope while SMASH uses a 4-m telescope, the photometric quality of SMASH is better, especially in dense star fields. The brighter photometric data may have been responsible for the identification of the above four stars as binaries by the work of \citet{2020ApJ...900..118N}. In addition, the MCPS $U$ photometry covers only 76.5\% of our sample, whereas the SMASH $u$ photometry covers 92.24\% of our sample. As shown in Figure \ref{nuegent_five}, the six stars should be single stars based on the SED fitting method, and they all show significant infrared excess indicating the existence of circumstellar dust.

\bsp	
\label{lastpage}
\end{document}